\documentclass{aa}
\usepackage{graphicx}
\usepackage{xcolor}
\usepackage{caption}
\usepackage{subcaption}
\usepackage{txfonts}
\usepackage{ulem}
\def\rsun{R$_{\odot}$}

\begin{document}

   \title{The Carina High-Contrast Imaging Project for massive Stars (CHIPS)}
   \subtitle{I. Methodology and proof of concept on QZ Car ($\equiv$ HD93206)}

   \author{A. Rainot\inst{1}, M. Reggiani\inst{1}, H. Sana\inst{1}, J. Bodensteiner\inst{1}, C. A. Gomez-Gonzalez\inst{2}, O.~Absil\inst{3}\fnmsep\thanks{F.R.S.-FNRS Research Associate}, V. Christiaens\inst{3,4,5}, P. Delorme\inst{6}, L. A. Almeida\inst{7,8}, S. Caballero-Nieves\inst{9}, J. De Ridder\inst{1}, K. Kratter\inst{10}, S. Lacour\inst{11}, J.-B. Le Bouquin\inst{6}, L. Pueyo\inst{12}, H. Zinnecker\inst{13}}
   \institute{Institute of Astronomy, KU Leuven, Celestijnlaan 200D, 3001 Leuven, Belgium\\
              \email{alan.rainot@kuleuven.be}
         \and
              Barcelona Supercomputing Center, carrer de John Maynard Keynes, 30, 08034 Barcelona, Spain
         \and
              Space sciences, Technologies and Astrophysics Research (STAR) Institute, Universit\'e de Li\`ege, 19 All\'ee du Six Ao\^ut, 4000 Li\`ege, Belgium
         \and
              Departamento de Astronom\'ia, Universidad de Chile, Casilla 36-D, Santiago, Chile
         \and
              School of Physics and Astronomy, Monash University, VIC 3800, Australia
         \and
              Universit\'e Grenoble Alpes, CNRS, IPAG, 38000 Grenoble, France
         \and
              Departamento de F\'isica, Universidade do Estado do Rio Grande do Norte, Mossor\'o, RN, Brazil
         \and
              Departamento de F\'isica Te\'orica e Experimental, Universidade Federal do Rio Grande do Norte, CP 1641, Natal, RN, 59072-970, Brazil
         \and
              Department of Aerospace Physics \& Space Sciences, Florida Institute of Technology 150 West University Blvd, Melbourne, FL 32901, USA
         \and
              Department of Astronomy, University of Arizona, Tucson, AZ 85721, USA
         \and
              LESIA, (UMR 8109), Observatoire de Paris, PSL, CNRS, UPMC, Universit\'e Paris-Diderot, 5 place Jules Janssen, 92195 Meudon, France
         \and
              Space Telescope Science Institute, 3700 San Martin Drive, Baltimore, MD, 21218, USA
         \and
              Universidad Autonoma de Chile, Avda Pedro de Valdivia 425, Providencia, Santiago de Chile, Chile
             }
   \date{Accepted}

   \authorrunning{A.~Rainot et al.}


  \abstract
   {Massive stars like company. However, low-mass  companions have remained extremely difficult to detect at angular separations ($\rho$) smaller than 1"  (approx.\ 1000-3000 au considering typical distance to nearby massive stars) given the large brightness contrast between the companion and the central star. Constraints on the low-mass end  of the companions mass-function for massive stars are however needed, for example to help distinguishing between various scenarios for the formation of massive stars.}
   {To obtain statistically significant constraint on the presence of low-mass companions beyond the typical detection limit of current surveys ($\Delta \mathrm{mag} \lesssim 5$ at $\rho \lesssim 1$"), we initiated a survey of O and Wolf-Rayet stars in the Carina region using the SPHERE coronagraphic instrument on the VLT. In this first paper, we aim to introduce the survey, to present the methodology and to demonstrate the capability of SPHERE for massive stars using the multiple system QZ~Car.}
   {We obtained VLT-SPHERE snapshot observations in the IRDIFS\_EXT mode, which combines the IFS and IRDIS sub-systems and simultaneously provides us four-dimension data cubes in two different field-of-view: 1.73" $\times$ 1.73" for IFS (39 spectral channels across the $YJH$ bands)  and 12" $\times$ 12" for IRDIS (two spectral channels across the $K$ band). Angular- and spectral-differential imaging techniques as well as PSF-fitting were applied to detect and measure the relative flux of the companions in each spectral channel. The latter  are then flux-calibrated using theoretical SED models of the central object and are compared to a grid of ATLAS9 atmosphere model and (pre-)main-sequence evolutionary tracks, providing a first estimate of the physical properties of the detected companions.}
   {Detection limits of 9~mag at $\rho > 200$~mas for IFS and as faint as 13~mag at $\rho > 1\farcs8$ for IRDIS (corresponding to sub-solar masses for potential companions) can be reached in snapshot observations of only a few minutes integration times, allowing us to detect 19 sources around the QZ~Car system. All but two are reported here for the first time. With near-IR magnitude contrasts in the range of 4 to 7.5~mag, the three brightest sources (Ab, Ad and E) are most likely physically bound, have masses in the range of 2 to 12~M$_\sun$ and are potentially co-eval with QZ~Car central system. The remaining sources  have flux contrast of $1.5\times10^5$ to $9.5 \times 10^6$ ($\Delta K \approx 11$ to 13~mag). Their presence can be explained by the local source density and they are thus probably chance alignments. If they are members of the Carina nebula, they would be sub-solar-mass pre-main sequence stars. }
   { Based on this proof of concept, we showed that VLT/SPHERE allows us to reach the sub-solar mass regime of the companion mass function. This paves the way for this type of observation with a large sample of massive stars to provide novel constraints on the multiplicity of massive stars in a region of the parameter space that has remained inaccessible so far.}
   \maketitle

\section{Introduction}\label{s:intro}

     The formation of massive stars remains one of the most important open questions in astronomy today \citep[e.g.,][]{zinnecker,tan}. Observing the early phases of massive stars formation  remains challenging at best: forming massive stars are rare and found at large distances, their formation timescale is short and they are born in an environment strongly obscured by gas and dust.

    Several formation scenarios have been proposed, among others: formation through stellar collisions and merging \citep{bonnel1998}, competitive accretion \citep{bonnel2001,bonnel2006}, monolithic collapse \citep{mckee,krumholz}. Except for the merger process, most theories agree on the need for dense and massive accretion disks to overcome the radiation barrier. These disks likely fragment under gravitational instabilities \citep{kratter} which may result in the formation of companions, however model predictions are still scarce. Studying the correlations between multiplicity characteristics may provide crucial observational constraints to distinguish between the different scenarios of massive star-formation and help in the development of future theoretical models.

     The multiplicity properties of massive stars have already been the subject of recent surveys in the Milky Way and nearby galaxies \citep[for a recent overview, see e.g.][]{sana2017}. Some studies focused on the spectroscopic analysis of young massive stellar clusters \citep{sana2012,almeida2017} and OB associations \citep{kobulnicky2012} and others on the high-angular astrometric observations of massive stars \citep{mason09, sana2014, aldoretta2015, gravity2018} in order to determine the binary fraction of massive stars in these regions.

     Among those previous studies, the Southern MAssive Stars at High angular resolution survey \citep[SMaSH+,][]{sana2014} was an ESO Very Large Telescope (VLT) Large Program (P89-P91) that combined optical interferometry (VLTI/PIONIER) and aperture masking (NACO/SAM) to search for mostly bright companions ($\Delta H<4$) in the angular separation regime $0.001\arcsec <\rho$\,$<$ $0.2$\arcsec\, around a large sample of O-type stars. The entire NACO field of view was further analysed to search for fainter ($\Delta H<8$) companions up to 8\arcsec.

     The SMaSH+ results showed the importance of such studies for the understanding of massive star formation. They concluded that almost all massive stars in their sample have at least one companion and that over 60\%\ have two or more. In addition,  a larger number of faint companions are seen at large separations, corresponding roughly to the outer edge of the accretion disk. This is in agreement with expectations from the theory of disk fragmentation. These companions may correspond to outward migrating clumps resulting from the fragmented accretion disk or from tidal capture. Investigating whether low-mass companions exist at closer separations or if there is a characteristic length at which the flux {\it vs.}\ separation distribution changes is therefore critical.

     Nevertheless, there remain large areas in the parameter space that have not been probed by these surveys, mostly due to instrumental limitations. In particular and while a rather complete view  of companions down to mass ratio of about 0.3 has now been achieved, the existing surveys have so-far failed to probe the lower-mass end of the companion mass-function. In the last few years, extreme Adaptive Optics (AO) instruments have come online, peering far deeper and more accurately than previously possible. Extreme AO, implemented at the VLT through the Spectro-Polarimetric High-contrast Exoplanet REsearch instrument \citep[SPHERE,][]{beuzit2019} provides the necessary spatial resolution and dynamics to search for faint companions to nearby massive stars.

    In this context, the Carina High-contrast Imaging Project of massive Stars (CHIPS) aims to characterise the immediate environment of a large sample of massive stars within 3\degr\ from $\eta$ Car. Ninety-three O- \& Wolf-Rayet type stars were selected from the Galactic O-Star Catalogue \citep[GOSC,][]{maiz2013} and the Galactic Wolf-Rayet catalogue \citep{crowther2015}. So far about half of the potential targets have been observed, which will be sufficient to obtain constraints on the occurrence rate of companions in the SPHERE separation range with a precision better than 7\%. SPHERE will allow us to investigate the presence and properties of massive star companions in the angular separation range of $0\farcs15$ to $5\farcs5$ (approx.\ 350-12,500 au) and $\Delta \mathrm{mag} \approx 12$ (mass-ratios > 0.03 on the main sequence). The range below a couple thousand au is particularly important as it corresponds to the approximated size of the accretion disk, where faint companions formed from the remnant of the fragmented disk could be found.

     The present paper is the first in a short series. Here, we aim to establish a proof-of-concept using the first VLT/SPHERE observations of the \object{QZ Car} multiple system.  QZ Car ($\equiv$~HD~93206) is a high-order multiple system composed of two spectroscopic binaries (Aa \& Ac) and three previously resolved companions within $7\arcsec$ (Ab, E \& B). The pair (Aa1,Aa2)  has a spectral type  O9.7~I~+~B2~V, and an orbital period of 20.7 days. The pair (Ac1,Ac2)  has a spectral type O8~III~+~O9~V, and a period of 6 days. These two binaries make up the central system (Aa,Ac), separated by roughly 30 milli-arcsec (mas) \citep{sana2014,sanchez2017}, and have a combined $H$ and $K_\mathrm{s}$-band magnitudes of 5.393 and 5.252, respectively. The companions Ab, E \& B were detected by the SMaSH+ survey at separations of $1\farcs00$, $2\farcs58$ and $7\farcs07$ from the central system, respectively.

     This paper is organized as follows. Sect.~\ref{s:data} presents the observations and data reduction. Sect.~\ref{s: analysis} describes the image post-processing algorithms as well as additional functionalities developed for our current studies. Results are discussed in Sect.~\ref{s: results} and our conclusions are presented in Sect.~\ref{s:Ccl}

\section{Observations and data reduction} \label{s:data}

  \subsection{Observations}\label{s:obs}

    The QZ~Car observations were obtained on Jan 25th, 2016 using the second generation VLT instrument SPHERE, situated on the Unit Telescope 3 at the Paranal observatory in Chile. SPHERE is a high-contrast imaging instrument combining an extreme adaptive optics system, coronagraphic masks and three different sub-systems with specific science goals. Our observations were executed in the IRDIFS extended mode (IRDIFS\_EXT) mode using the Integral Field Spectrograph \citep[IFS,][]{claudi2008} and the Infra-Red Dual-beam Imaging and Spectroscopy \citep[IRDIS,][]{dolhen2008} sub-systems.

    \begin{table}
    \caption{Observing setup and atmospheric conditions for QZ Car's {\sc flux} (F) and {\sc object} (O) observations.}
    \label{table:Obs}
    \centering
    \begin{tabular}{l c c }
    \hline\hline
    \vspace*{-3mm}\\
    Instrument & IFS & IRDIS \\
    \hline
    Number of DITs (NDIT) (O) & 16  & 4     \\
     Detector Integration Time (DIT) (O) [s]  & 4   & 4     \\
    Number of DITs (NDIT) (F)   & 4  & 4    \\
     Detector Integration Time (DIT) (F) [s]   & 16   & 8     \\
    Neutral Density Filter     & --   & ND\_2 \\
    Airmass    & 1.3   &    1.3   \\
    Parallactic Angle variation (\degr)  & 3.4 &   3.5    \\
    Seeing at zenith     &   0.9 &    0.9   \\
    Average Coherence time $\tau_{0}$ (ms) & 3 &     3 \\
    \hline
    \end{tabular}
    \end{table}

     IFS images have a size of 290 $\times$ 290 pixels and a pixel size of 7.4~mas, hence corresponding to a field-of-view (FoV) of 1\farcs73\ $\times$ 1\farcs73 \ on the sky. The IRDIS camera has 1024 $\times$ 1024 pixels, covering a 12\arcsec\ $\times$ 12\arcsec\ FoV with a pixel size of 12.25 mas.
    The IRDIFS\_EXT mode was chosen to allow combining the $YJH$-band observations with IFS to dual-band $K$-band observations with IRDIS. With its small FoV and spectroscopic capabilities, IFS allowed us to both detect and characterise companions at short separations. The larger FoV of IRDIS provided additional information on the local density of faint objects.

    The observation sequence was composed of three types of observations: (i) {\sc centre (C)}, allowing us to compute the centroid location of the coronagraph; (ii) {\sc flux (F)}, to obtain a reference flux point-spread function (PSF) of the central objects; and (iii) {\sc object (O)}, with the central star blocked by the coronagraph, hence delivering the scientific images that will be scrutinised to search for faint, nearby companions. {\sc flux} observations were performed with the central star outside the coronagraph. The F-C-O sequence was repeated three times.
    Due to the brightness of QZ Car ($ H < 5.5$), we used the neutral density filter ND2.0 -- delivering a transmission of the order of $10^{-2}$ -- for the F and C observations of both instruments. The telescope was set to pupil tracking, i.e. the centroid of the field is fixed on the science object and the sky rotates around it, as required for the later post-processing algorithm which uses the angular information of the movement of companions on an image, or angular differential imaging \citep{marois06}.

    For the IFS {\sc object} observations, we used 16 DITs of $4$~s, for a total exposure time of $\sim$64~s. With IRDIS we chose 4 DITs of $4$~s, giving an integration time of $16$~s. Similarly we adopted NDIT $\times$ DIT of 4 $\times$ 16~s and NDIT $\times$ DIT of 4 $\times$ 8~s for the {\sc flux} exposures with IFS and IRDIS, respectively. The observing setup and  atmospheric  conditions are detailed in Table \ref{table:Obs}. As a result of the {\sc object} observations, we have obtained four-dimensional (4D) IFS and IRDIS data cubes of the QZ~Car system. The IFS cubes are  composed of spatial (2D) images of the IFS FoV for each of the 39 wavelengths channels (from 0.9 to 1.6 $\mu$m) and 48 sky rotations (due to the pupil tracking). The IRDIS data cubes contain 2D pixel images at each of the two wavelengths channels ($K_1$ and $K_2$) and 48 sky rotations, covering a total parallactic angle variation of about 3.5\degr.

  \subsection{Science data reduction} \label{s: reduction}

  	The data reduction of IRDIS and IFS images was processed by the SPHERE Data centre  \citep[DC]{delorme2017} at the Institut de Planetologie et d'Astrophysique de Grenoble (IPAG)\footnote{\url{http://ipag.osug.fr/?lang=en}}. The SPHERE-DC process is standardised in terms of astronomical data reduction: removing bad pixels, dark and flat frames and estimating the bias in each exposure. They also calibrate the astrometry associated to the science frames using the on-sky calibrations from \citet{maire2016}, i.e. a True North correction value of $1.75\pm0.08^\circ$ and a plate scale of $7.46\pm0.02$ mas/pixel for IFS and $12.255\pm0.009$ mas/pixel for IRDIS. The system uses a modified version of the SPHERE Esorex pipeline\footnote{\url{http://eso.org/sci/software/pipelines}} that is functional, automated and can be accessed by the user if requested. The end result from this data reduction is the reduced 4D science data cube, tables containing the wavelengths and rotational angles, and the 3D PSF cubes (see~Sect.~\ref{s:psf}).

  \subsection{The point spread function of QZ~Car} \label{s:psf}

    The {\sc flux} observations are images of the central star taken without the coronagraph to estimate the PSF and flux of the central star. Three such PSF observations were taken during our observing sequence and delivered data cubes that contained the 2D images of the field at each of the instruments' respective wavelength channels. To increase the signal-to-noise ratio ($S/N$) of the PSF frames and to average out the observing conditions, the median of the three PSF frames at each wavelength was computed so only a single 3D data cube was left. The obtained PSF is to be used for the companion modelling and characterisation techniques introduced in Sect.~\ref{s: analysis}. However, in the case of QZ~Car, additional complication arose in the IFS PSF frames.

    The pair of close binaries at the core of QZ Car's multiple system is separated by roughly 30~mas. This is just about the diffraction limit of an 8.2~m telescope in the $Y$ to $H$ band and sufficient for QZ~Car's point spread function (PSF) in IFS to display an elongated shape in the {\sc flux} images. This had unintended consequences in the data processing as the reference PSF obtained from the IFS {\sc flux} images is used to create a normalised PSF needed to inject artificial companions at different stages in the analysis, hence propagating the PSF's deformation and leading to a number of artefacts. Therefore we adopted another reference PSF from an IFS observation of HD~93129A taken on the night of February 10, 2016. While HD~93129A is itself a long period binary, it was unresolved at the time of the observations \citep{maiz2017}. The flux of HD~93129A's PSF is not the same as the original PSF of QZ Car. It was therefore scaled so that the new reference IFS PSF images have the same integrated flux as the QZ~Car IFS PSF frames. This allows us to retain the original flux information while adopting a more representative PSF shape.

    Error estimation of the total flux measured from the PSF was accomplished by computing the standard deviation of the flux for all the {\sc flux} images.

\section{Data Analysis} \label{s: analysis}

  \begin{figure*}
     \centering
        \includegraphics[width=.95\columnwidth]{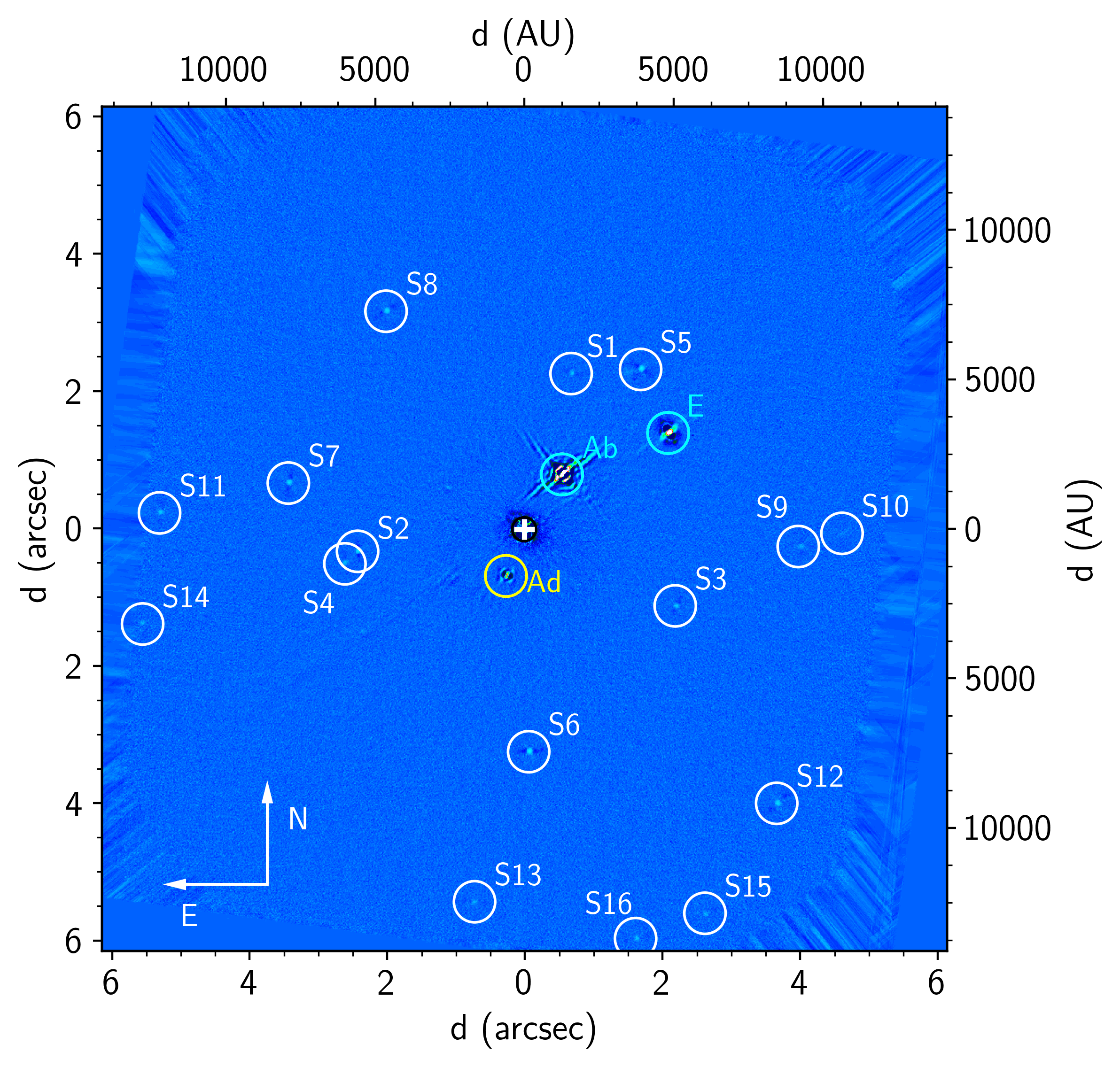}\hspace*{5mm}
        \includegraphics[width=.96\columnwidth]{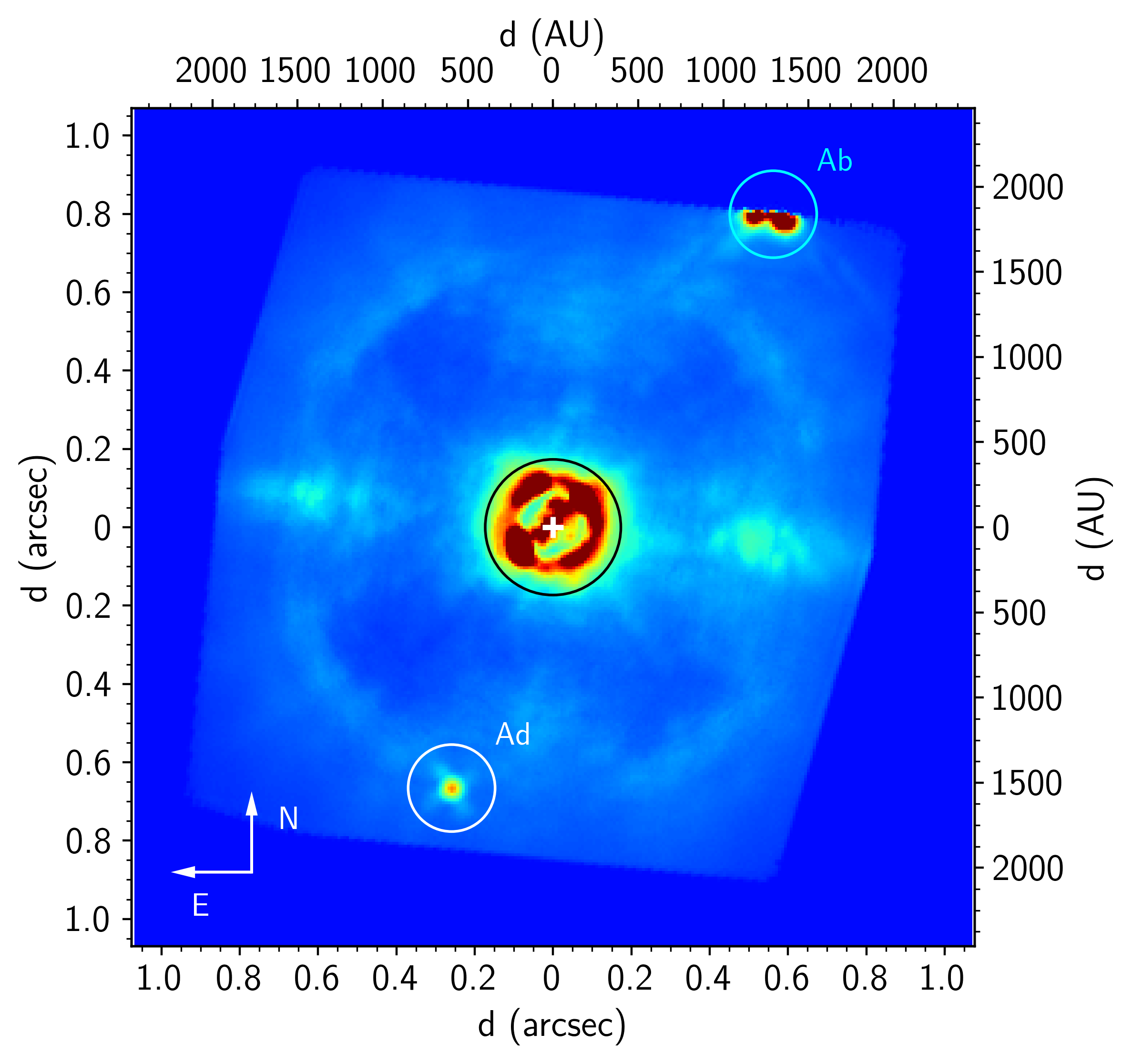}
      \caption{Post-processed science frames from IRDIS (left) and IFS (right) instruments. Angular and spectral differential imaging was applied with one principal component used for modelling and subtraction of the residual stellar light (see Sect.~\ref{s: analysis}). The IRDIS image combines the two wavelengths channels K1 and K2. Dark circles on the images show the $0\farcs173$-diameter size of the coronagraph. The blue circles indicate the previously resolved companions Ab and E. White circles indicate the IRDIS sources detected at the $5\sigma$ level while yellow circles indicate the new companion QZ~Car Ad. The IFS image is $1\farcs73 \times 1\farcs73$; the IRDIS is $12\arcsec \times 12\arcsec$.}
      \label{f: fov}
   \end{figure*}

	Once the data were reduced, image post-processing algorithms for high-contrast imaging were used on the target frames. In this study, we made use of the Vortex Imaging Processing package\footnote{\url{https://github.com/vortex-exoplanet/VIP}} \citep[VIP,][]{carlos2017} and a PSF-fitting technique \citep[][]{bodensteiner19}. VIP is an open-source python package for high-contrast imaging data processing that is instrument-agnostic. It was developed for exoplanet research, disk detection and characterization. It is able to perform Angular, Reference and Spectral Differential Imaging (ADI, RDI, SDI respectively) and ADI-SDI simultaneously based on matrix approximation with Principal Component Analysis \citep[][]{amara2012,soummer2012}. We contributed to the implementation of 4D data analysis and IFS support.

    For comparison, a PSF fitting approach was also used to obtain the photometry of sources on IRDIS images and on companions which VIP could not characterise because too close to the edges of the detector (see section \ref{s: PSF-fit}).

  Post-processed PCA/ADI images for the IRDIS and IFS science cubes are presented on Fig.~\ref{f: fov}. From the IRDIS image, previously known companions Ab and E are clearly distinguishable. However, these are not the only detected companions as over a dozen of other point sources are now clearly visible on the image. The IFS image gives a close-up view of the system. The Ab companion is still present at the top of the image though cropped and cannot be analysed. On the South of the IFS image, a previously unknown companion (Ad) is clearly visible and delivering enough flux for a first spectral analysis (see Sect.~\ref{s: SED}).

 \begin{table*}
    \caption{Angular separations ($\rho$) and projected physical separations ($d$), position angles (PA, measured from North to East), magnitude contrasts in the $K_1$ and $K_2$ bands ($\Delta K_{1,2}$) and spurious alignment probabilities ($P_\mathrm{spur}$) for each detected IRDIS sources. Parameters in \textit{italic} were estimated using the PSF-fitting method described in Sect.~\ref{s: PSF-fit}. All others are from the Simplex-MC method.
   }
    \label{table:2}
    \centering
    \begin{tabular}{l c c c c c }     
        \hline\hline
       Source & Ad & Ab & E & S1 & S2  \\
       \hline

       $\rho$ (mas) & 729.1 $\pm$ 1.4 & 1002.9 $\pm$ 1.7 & 2590.4 $\pm$ 4.4 & 2429.4 $\pm$ 6.9 & 2475.8 $\pm$ 5.9 \\
       $d$ ($10^3$~au) &
1.7 $\pm$ 0.1 &
2.3 $\pm$ 0.1 &
5.9 $\pm$ 0.1 &
5.6 $\pm$ 0.1 &
5.7 $\pm$ 0.1 \\

       PA ($^\circ$) & 169.9 $\pm$ 0.1 & 335.6 $\pm$ 0.1 & 314.3 $\pm$ 0.1 & 343.1 $\pm$ 0.1 & 197.8 $\pm$ 0.1 \\
       $\Delta K_{1}$ (mag) & 7.5 $\pm$ 0.1 & 4.4 $\pm$ 0.1 & 7.2 $\pm$ 0.1 & 12.0  $\pm$ 0.1 & 11.2 $\pm$ 0.1 \\
       $\Delta K_{2}$ (mag) & 7.6 $\pm$ 0.1 & 4.4 $\pm$ 0.1 & 7.1 $\pm$ 0.1 & 12.2 $\pm$ 0.2 & 11.1 $\pm$ 0.1 \\

       $P_\mathrm{spur}$  (\%) & 0.2 & 0.1 & 1.5 & 39.0  & 26 \\

    \hline
\vspace*{3mm}\\
    \hline\hline
       Source & S3 & S4 & S5 & S6 & S7\\
       \hline

       $\rho$ (mas) & 2470.8 $\pm$ 6.8 & 2661.4 $\pm$ 7.5 & 2955.7 $\pm$ 5.7 & 3298.6 $\pm$ 5.9 & 3553.9 $\pm$ 6.6 \\
       $d$ ($10^3$~au) &
 5.8 $\pm$ 0.1 &
 6.2 $\pm$ 0.1 &
 $6.8\pm0.1$ &
 $7.6\pm0.1$ &
 $8.2\pm0.1$\\

       PA ($^\circ$) & 205.9 $\pm$ 0.4 & 221.8 $\pm$ 0.4 & 334.3 $\pm$ 0.1 & 191.8 $\pm$ 0.1 & 89.1 $\pm$ 0.1\\
       $\Delta K_{1}$ (mag) & 11.8 $\pm$ 0.1 & 12.2 $\pm$ 0.1 & 11.1 $\pm$ 0.1 & 10.9 $\pm$ 0.1 & 11.5 $\pm$ 0.1\\
       $\Delta K_{2}$ (mag) & 11.4 $\pm$ 0.1 & 11.8 $\pm$ 0.1 & 10.9 $\pm$ 0.1 & 10.9 $\pm$ 0.1 & 11.3 $\pm$ 0.1\\
       $P_\mathrm{spur}$  (\%) & 33.0 & 44.0 & 33.0 & 36.0 & 52.0 \\

    \hline
\vspace*{3mm}\\
    \hline \hline
       Source & S8 & S9 & S10 & S11 & S12\\
       \hline
       $\rho$ (mas) & 3836.2 $\pm$ 7.6 & 4114.0 $\pm$ 10.0 & 4722.8 $\pm$  11 & 5401.2 $\pm$ 11.0 & 5537.3 $\pm$ 10.0 \\
       $d$ ($10^3$~au) & 8.8 $\pm$ 0.1 & 9.5 $\pm$ 0.1 & 10.9 $\pm$ 0.1 & 12.4 $\pm$ 0.1 & 12.7 $\pm$ 0.1\\

       PA ($^\circ$) & 42.5 $\pm$ 0.1 & 266.5 $\pm$ 0.1 & 269.6 $\pm$ 0.1 & 87.3 $\pm$ 0.1 & 222.8 $\pm$ 0.1 \\
       $\Delta K_{1}$ (mag) & 11.5 $\pm$ 0.1 & 12.5 $\pm$ 0.1 & 13.0 $\pm$ 0.2 & 11.9 $\pm$ 0.1 & 11.4 $\pm$ 0.1 \\
       $\Delta K_{2}$ (mag) & 11.5 $\pm$ 0.1 & 11.9 $\pm$ 0.1 & 13.1 $\pm$ 0.3 & 12.1 $\pm$ 0.1 & 11.1 $\pm$ 0.1\\
       $P_\mathrm{spur}$ (\%)  &  60.0 & 78.0 & 92.0 & 90.0  & 81.0 \\

    \hline
    \vspace*{3mm}\\
        \hline \hline
       Source & S13 & S14 & S15 & S16 \\
       \hline
       $\rho$ (mas) & 5611.2 $\pm$ 10.1 & 5866.5 $\pm$ 10.3 & \textit{6313.1 $\pm$ 11.0} & \textit{6327.1 $\pm$ 11.0} \\
       $d$ ($10^3$~au) & 12.9 $\pm$ 0.1 & 13.5 $\pm$ 0.1 & \textit{14.5 $\pm$ 0.1} & \textit{14.5 $\pm$ 0.1}\\

       PA ($^\circ$) & 172.2 $\pm$ 0.1 & 103.9 $\pm$ 0.1 & \textit{164.9 $\pm$ 0.1} & \textit{155.1 $\pm$ 0.1} \\
       $\Delta K_{1}$ & 12.5 $\pm$ 0.1 & 12.5 $\pm$ 0.1 & \textit{12.4 $\pm$ 0.1} & \textit{12.2 $\pm$ 0.1} \\
       $\Delta K_{2}$ & 12.1 $\pm$ 0.2 & 12.3 $\pm$ 0.2 & \textit{12.1 $\pm$ 0.1} & \textit{12.1 $\pm$ 0.1} \\

       $P_\mathrm{spur}$ (\%) & 94.0 & 96.0 & 96.0 & 95.0 & \\

    \hline

    \end{tabular}
    \end{table*}

  \subsection{Companion Detection} \label{s: detect}

    A visual inspection of the final IRDIS and IFS PCA images displayed in Fig.~\ref{f: fov} reveals a handful of rather bright companions and a larger number of much fainter point sources.
    To evaluate which ones are true detections, we first estimated their signal-to-noise ratio ($S/N$) using the $S/N$ map function implemented in VIP.
     This module computes the $S/N$ at every pixel of the frame as defined in \citet{mawet2014}. From this map we set our detection limit to $S/N = 5$. As expected, companions Ab and E as well as the new companion Ad (detected in both IFS and IRDIS) have large $S/N$. 16 other companion candidates are at separations beyond the IFS FoV and are detected in the IRDIS image with $S/N > 5$. They are marked \textit{S1} to \textit{S16} in Fig.~\ref{f: fov}, yielding a total of 19 individual sources detected within $6\farcs2$ from the QZ~Car central quadruple system.

  \subsection{Source characterisation} \label{s: charact}

    Once we identified the true sources, we retrieved their position and contrast with respect to the central star. Starting from guess positions estimated from the post-processed frames, we measured accurate angular separations ($\rho$), position angles (PA) and flux contrast in each wavelength channel and for each companion with three different methods described below. Two methods included in the VIP package were used for Ad, Ab and E companions and PSF fitting for all S sources. Final results are provided in Table ~\ref{table:2} and App.~\ref{a: Adspec} for IRDIS and IFS respectively.

    \subsubsection{VIP} \label{s: VIP}

    We first measured the flux of the stars using aperture photometry spectral channel by spectral channel, which provided us with a first estimate of the spectrum. This initial guess was then passed onto the Simplex Nelder-Mead optimisation (hereafter referred to as the Simplex method) of VIP which estimated position and flux parameters by applying a NEGative Fake Companion technique (NEGFC). This method consists in inserting negative artificial sources in the individual frames, varying at the same time their brightness and position (starting from the values measured by aperture photometry). The artificial companions are obtained from the unsaturated PSF of the central star, measured in the {\sc flux} observations. The residuals in the final images are then computed and compared to the background noise, measured in an annulus at the same radial distance. The combination of brightness and location that minimizes the residuals are estimated through a Nelder-Mead minimization algorithm. This provides reference fluxes for each spectral channel. Dividing the fluxes of the different companions by the reference fluxes from the PSF data cubes, we obtained flux or magnitude contrasts at each wavelength channel.

Results from this simplex method were then injected into the MCMC engine (also available in VIP) to estimate confidence intervals for the angular separations, PAs and fluxes for each wavelength channel and each target. In this way we obtained an (uncalibrated) IFS low-resolution spectrum for companion Ad and two flux values for all IRDIS sources. For sources with small radial distance difference between each other ($\Delta \rho < 200$ mas), masks were applied on sources other than the target. This was to prevent issues with parameter estimation when multiple sources are in the same annulus when applying the simplex optimisation routine. For this purpose, circular masks were taken at the same radial distance as the source we wished to mask, but at a different position angle to preserve the noise properties. A rotation was then applied on the mask in order to preserve the radial dependence of the noise. For Ab, the MCMC algorithm failed to retrieve the $K_{1}$ uncertainties. The MCMC failed to converge for S13 to S16 while nor Simplex nor MCMC could  be used for S15 and S16. In either case, this is caused by the fact that  the field-of-view is too small to contain the full annulus at the  target separation, i.e. the sources are too close to the edge of the field.

        \begin{figure*}
    \centering
    \begin{subfigure}{.5\textwidth}
      \centering
       \includegraphics[scale=0.31]{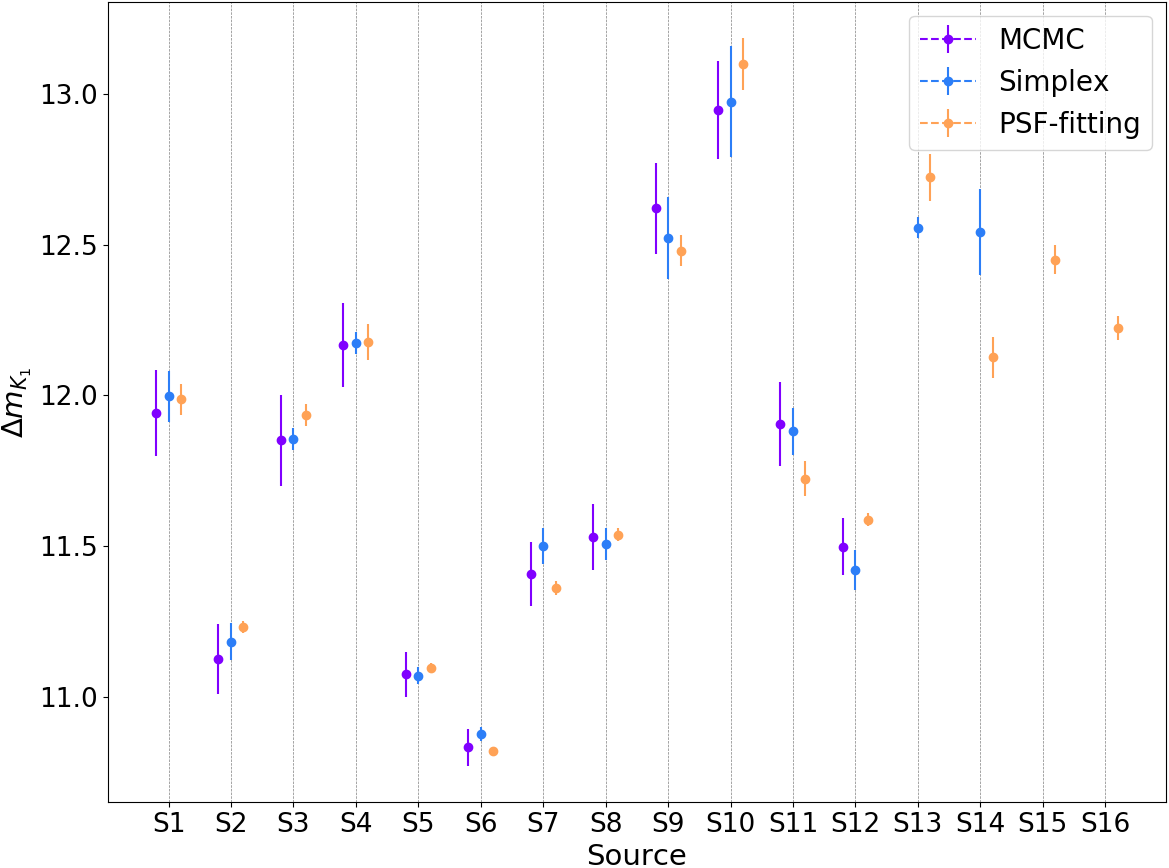}
     \end{subfigure}%
    \begin{subfigure}{.5\textwidth}
      \centering
       \includegraphics[scale=0.31]{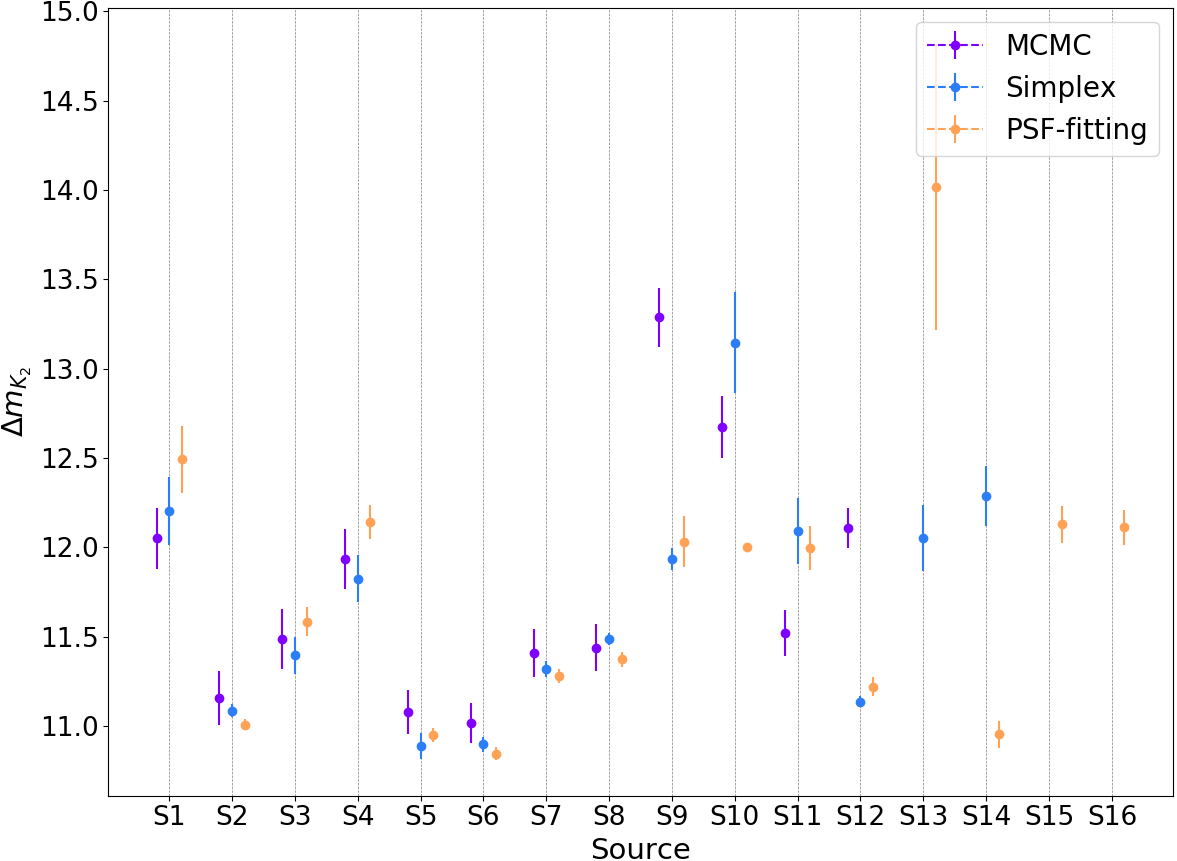}
     \end{subfigure}
     \caption{ \textit{Left :} $K_1$. \textit{Right :} $K_2$. IRDIS contrast magnitudes and associated errors of S sources obtained with three different methods: MCMC, Simplex-MC and PSF-fitting. Some errors for Simplex-MC and PSF fitting are small and not seen on these plots.}
     \label{f: dmag}
     \end{figure*}

        \begin{table*}
        \caption{Parameters of the QZ~Car central quadruple system adopted for the FASTWIND modelling of the system's SED. Spectral types are taken from \citet{sanchez2017} with calibration tables from \citet{martins05}.}
        \label{table:1}
        \centering
        \begin{tabular}{c c c c c c c c c }
        \hline\hline
        \vspace*{-3mm}\\
        Component & Spectral Type & $T_{\mathrm{eff}}$ & $\log g$ & $R_{\ast}$ & $M_{\ast}$ & $\log L_{\ast}$ & $\log\dot{M}$ & $v_{\infty}$\\
         & & (K) &  & (R$_{\odot}$) & (M$_{\odot}$) & [L$_{\odot}$] &[M$_{\odot}$\,yr$^{-1}$]      & (km\,s$^{-1}$)\\
        \hline
           Aa1 & O9.7~I & 30463 & 3.2 & 22.1 & 27.4   & 5.6 &$-5.7$ & 1794\\
           Aa2 & B2~V   & 20000 & 4.3  &  3.0 & 6.5   & 3.1 & $-9.7$ & 1186\\
           Ac1 & O8~III & 33961 & 3.6 & 13.7 & 25.4 & 5.3 & $-6.2$ & 2191\\
           Ac2 & O9~V   & 32882 & 3.92 &  7.53 & 17.1   & 4.7 & $-7.3$ & 2427\\
        \hline
        \end{tabular}
        \end{table*}

    In order to test the accuracy of the MCMC confidence intervals for IRDIS sources, we further implemented a Monte Carlo (MC) method. Firstly, all detected sources are masked. Secondly, a number (25 for our case) of artificial sources are injected at the same radial distance and with the same flux as a given companion, but at varying position angles. The flux and position of the artificial companions are then measured using the Simplex algorithm and compared to the input values. The corresponding standard deviations finally yield  an estimate of the  1$\sigma$ error on the flux and positions of the considered companion. The process in then repeated for all sources. The results will be discussed in the next section.

The previous methods only take into account the statistical uncertainties from the image processing. For the final photometric errors we took into account the flux variations of the unsaturated images of the central star described in Sect.~\ref{s:psf}. For the calculation of the astrometric uncertainties, we adopted the plate scale and astrometric calibration precision given by \cite{maire2016} and the ESO SPHERE user manual. The final astrometric errors are obtained by a quadratic sum of the Simplex-MC measurement errors, the star centre position uncertainty \citep[1.2 mas, from][]{zurlo2016}, the plate scale precision of 0.021 mas/pix for IRDIS and 0.02 mas/pix for IFS, the true north uncertainty ($\pm 0.08\deg$), and the dithering procedure accuracy \citep[0.74 mas,][]{zurlo2016}.

    \subsubsection{PSF fitting} \label{s: PSF-fit}

    For all IRDIS candidates beyond 2\arcsec, the central star's influence is limited and the background noise dominates (see Sect~ \ref{s: limits}). Therefore the use of ADI and SDI techniques is not necessarily needed to derive precise astrometry and photometry. A more widely used strategy in astronomical imaging is to use a PSF-fitting technique which provides accurate position and flux values for the S sources (Ad, Ab \& E $< 2\farcs$). Our PSF-fitting method is based on the \texttt{photutils}\footnote{\url{https://photutils.readthedocs.io}} python package along with an effective PSF model developed by \citet{anderson2000} and is described in \citet{bodensteiner19}.

   For this, we used the derotated and collapsed images which maximise the S/N in both $K_1$ and $K_2$. The {\sc flux} observations with IRDIS are used to establish an accurate PSF model. This is then fitted to each source individually in order to obtain accurate positions and flux estimates. This technique is very useful for sources that are detected in the collapsed images but are too close to the edges of the frames for ADI techniques, i.e.\ S15 and S16. However, in the $K_2$ band, PSF fitting could not converge to a good solution for sources S10, S13 and S14 as they are too close to the detection limit and do strongly benefit from the ADI post-processing.

    A comparison plot of the magnitude contrasts obtained between the three methods used in this paper is shown in Fig.~\ref{f: dmag} for the $K$  bands. A comparison between the $X,Y$ coordinates can be found in App. \ref{a: comparison}. These figures show that the positions and (in most cases) contrast values from the three methods are in excellent agreement, for sources S1 to S8, beyond which discrepancies arise. Sources too close to the edge could not be fit with Simplex (S15-S16) nor MCMC (S13-S16), while the PSF-fitting approach failed for the faintest object in the $K_2$-band (S10, S13, S14). Aside from these differences in the best-fit values, MCMC usually yields magnitude contrast errors that are a factor of few larger than those obtained with the Simplex+MC approach and with PSF-fitting. It is likely that the true errors lay somewhat in the middle. From the model atmosphere-fit in the next section, we have strong evidence that the MCMC contrast errors are likely overestimated by a factor of three to four while the Simplex+MC  and PSF-fitting uncertainties are too small by a factor of about two. The order of magnitudes is however correct. For future work and given the fact that the MCMC is much more computationally intensive, the present comparison certainly favours the use of PSF-fitting for most sources, but the one that are the closest to the detection limit. Final results adopted from the Simplex+MC or from the PSF fitting techniques are given in Table~\ref{table:2}.

      \begin{figure}
      \centering
      \includegraphics[width=\hsize]{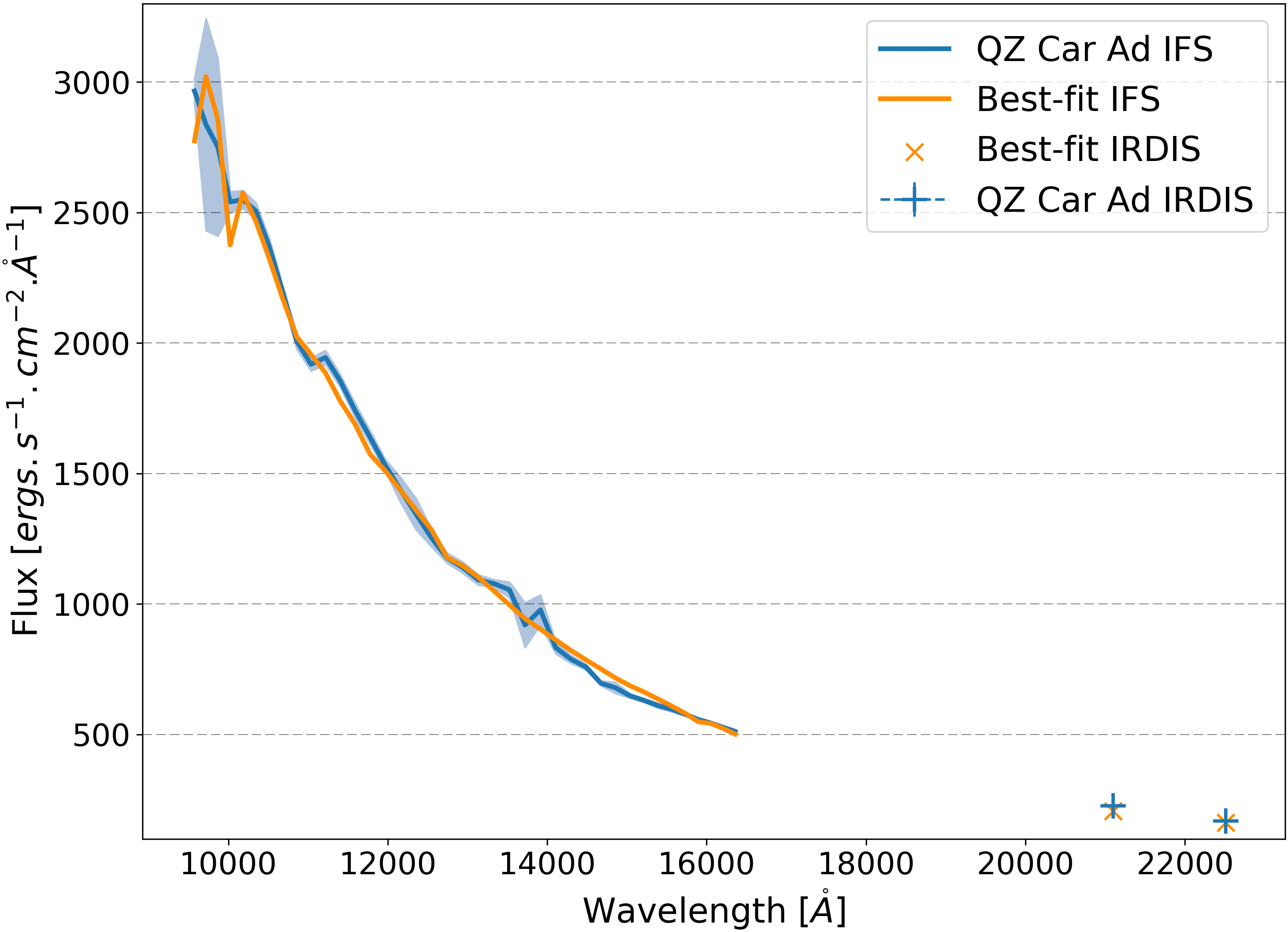}
         \caption{IFS+IRDIS flux-calibrated spectrum of QZ Car Ad at a reference distance of 100~R$_\odot$. The plain line gives the best-fit ATLAS9 model with $T_\mathrm{eff}=8896$~K, $\log g=4.27$ and $R=1.72$~R$_\odot$. Shaded area represents the 1-$\sigma$ uncertainties on the observed spectra.}
      \label{f: sed}
      \end{figure}

    \subsection{Flux calibration} \label{s: flux-cal}

    To obtain the absolute fluxes of the companions, we would require a flux calibrated spectrum of the central QZ Car system in the same wavelength range as that of our SPHERE observations ($Y$ to $K$). Unfortunately no such spectrum is available. To circumvent this issue, we modelled the spectral energy distribution of QZ~Car's central quadruple system using the non-local thermodynamic equilibrium (non-LTE) atmosphere code FASTWIND \citep[][]{puls2005,rivero2011}. Each component of the QZ Car's central system was modelled separately and their contribution within the PSF then combined. The parameters for the computation were obtained using the spectral types from \citet{sanchez2017} and the corresponding O-star calibration tables from \citet{martins05}. Parameters for the Aa2 component were found using a combination from \citet{trundle2007} and \citet{parkin2011}. We also calculated the mass-loss rate ($\dot{M}$) and terminal wind velocities ($v_\infty$) for each stellar component following \citet{vink2001} as these are needed input for FASTWIND. Results are summarised in Table \ref{table:1}.

    Once spectra for all the four central components were calculated and combined, we multiplied the contrast fluxes calculated previously by the model spectrum of QZ Car to obtain the absolute fluxes of each companion in the different wavelength bands. In particular, the IFS+IRDIS flux-calibrated spectrum of the Ad companion is displayed in Fig.~\ref{f: sed}. Throughout this process and later on in Sect.~\ref{s: results}, one needed to adopt a reference radius for the sphere at the surface of which the flux is computed. Without loss of generality, we arbitrarily adopted a value of 100~\rsun. We emphasize that this value has no physical meaning.

\begin{figure*}
\centering
\begin{subfigure}{.5\textwidth}
  \centering
  \includegraphics[scale=0.33]{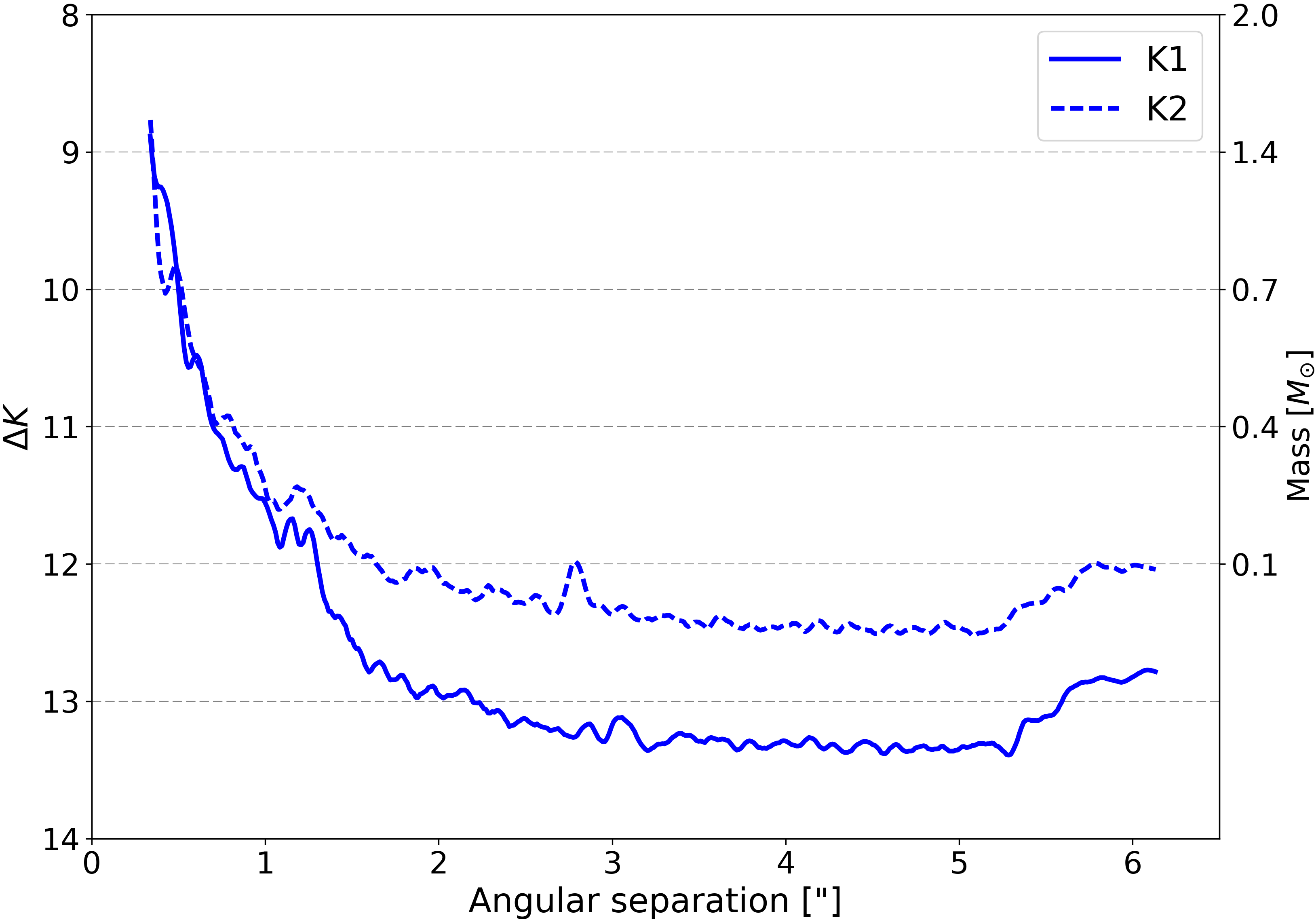}
 \end{subfigure}%
\begin{subfigure}{.5\textwidth}
  \centering
  \includegraphics[scale=0.33]{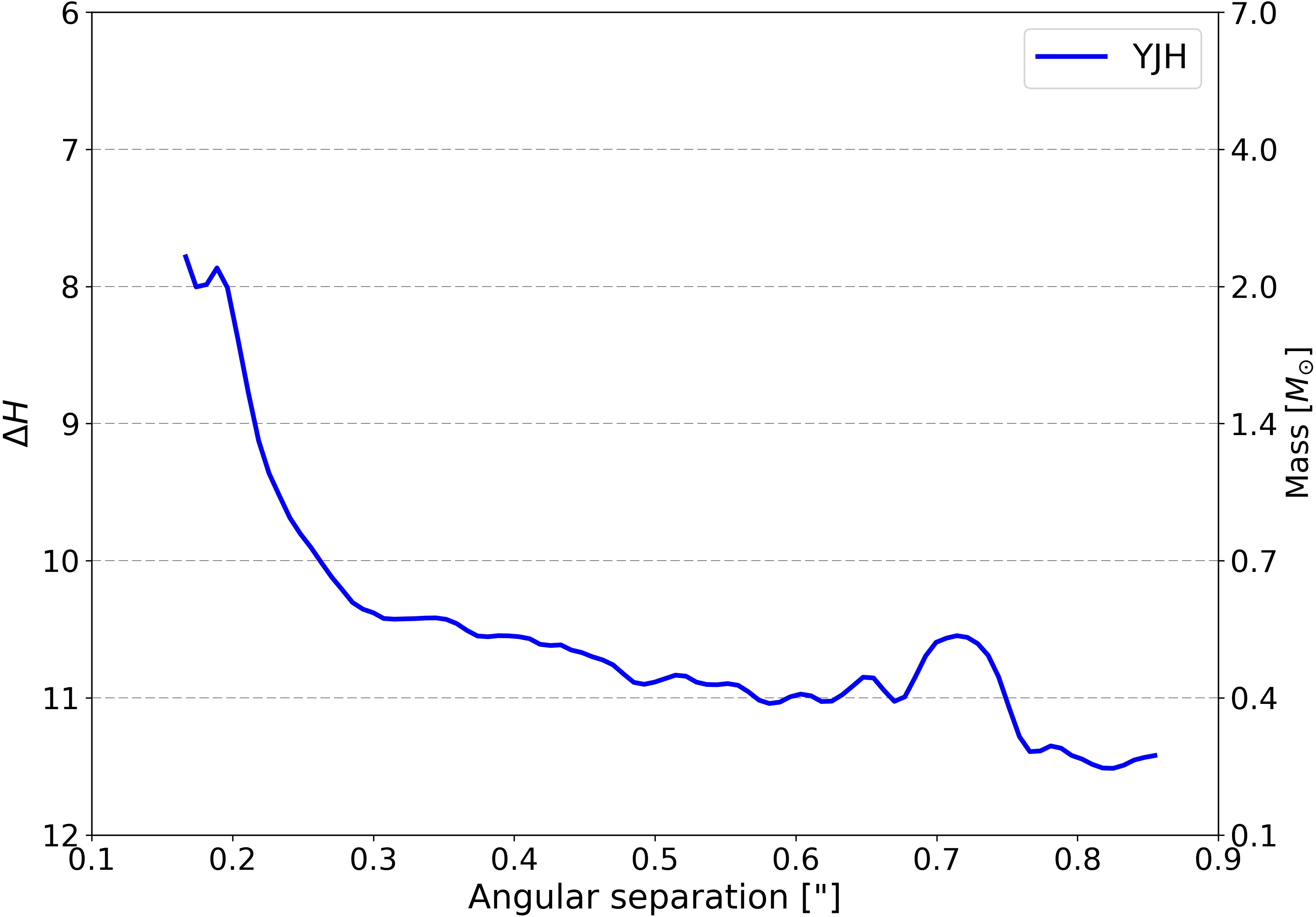}
 \end{subfigure}
 \caption{IRDIS (left) and IFS (right) detection limits (blue). The contrasts are given in the H-band for IFS and a mass scale is also provided on the right-hand axis. These masses were estimated using pre-main sequence and main-sequence tracks from \citet{siess2000}. Bumps at $\sim 0\farcs$7 in the IFS contrast curves results from the trace of the deformable mirror of the instrument at this radial distance.}
 \label{f: contrast}
 \end{figure*}

    \subsection{Detection limits} \label{s: limits}

      In this section, we estimate the sensitivity of our observations in terms of magnitude difference as a function of the angular separation $\rho$ to the central object. Using the VIP contrast curve modules we computed the contrast limits for a chosen $\sigma$ level by injecting artificial stars (based on the scaled PSF of HD~93129A, see Sect.~\ref{s:psf}) and calculating the noise at different radial distances from the centre. This implementation takes into account the small sample statistics correction proposed in \citet{mawet2014}. In order to avoid interference from the bright companions, all sources were masked (see Sect.~\ref{s: VIP}).
      Although this significantly increases the quality of the contrast curves, small artefacts with a 0.2~mag amplitude remain visible in the contrast curves at a radial separation of 3\arcsec.
    The 5-$\sigma$ sensitivity curves that we obtained are presented in Fig.~\ref{f: contrast}.

    A contrast better than 8~mag is achieved at 200~mas with IFS, and as large as 11~mag at $\rho > 600$~mas. These magnitude differences correspond to flux contrasts of $1.5\times10^{-4}$ and $4\times10^{-5}$, respectively.  Such detection limits are in line with past SPHERE observations \citep{zurlo2016,mesa2019} in  IRDIFS\_EXT mode, if we consider that our total exposure time was roughly a minute with both IFS and IRDIS. Using a mass scale from \citet{siess2000}'s evolutionary tracks, stars with masses $ < 1 M_{\odot}$ could be easily detected by this system. Similarly, IRDIS delivers contrast better than 9 mag at $0\farcs4$ and of better than 13~mag at separations larger than $2\farcs0$. This is about 5 magnitudes deeper than previously achieved with SMaSH+, and up to 8~mag deeper in the poorly mapped region around 400~mas demonstrating the complementing capabilities of SPHERE with respect to previously obtained high-angular resolution observations of massive stars.

  \begin{figure}
  \centering
  \includegraphics[width=\hsize]{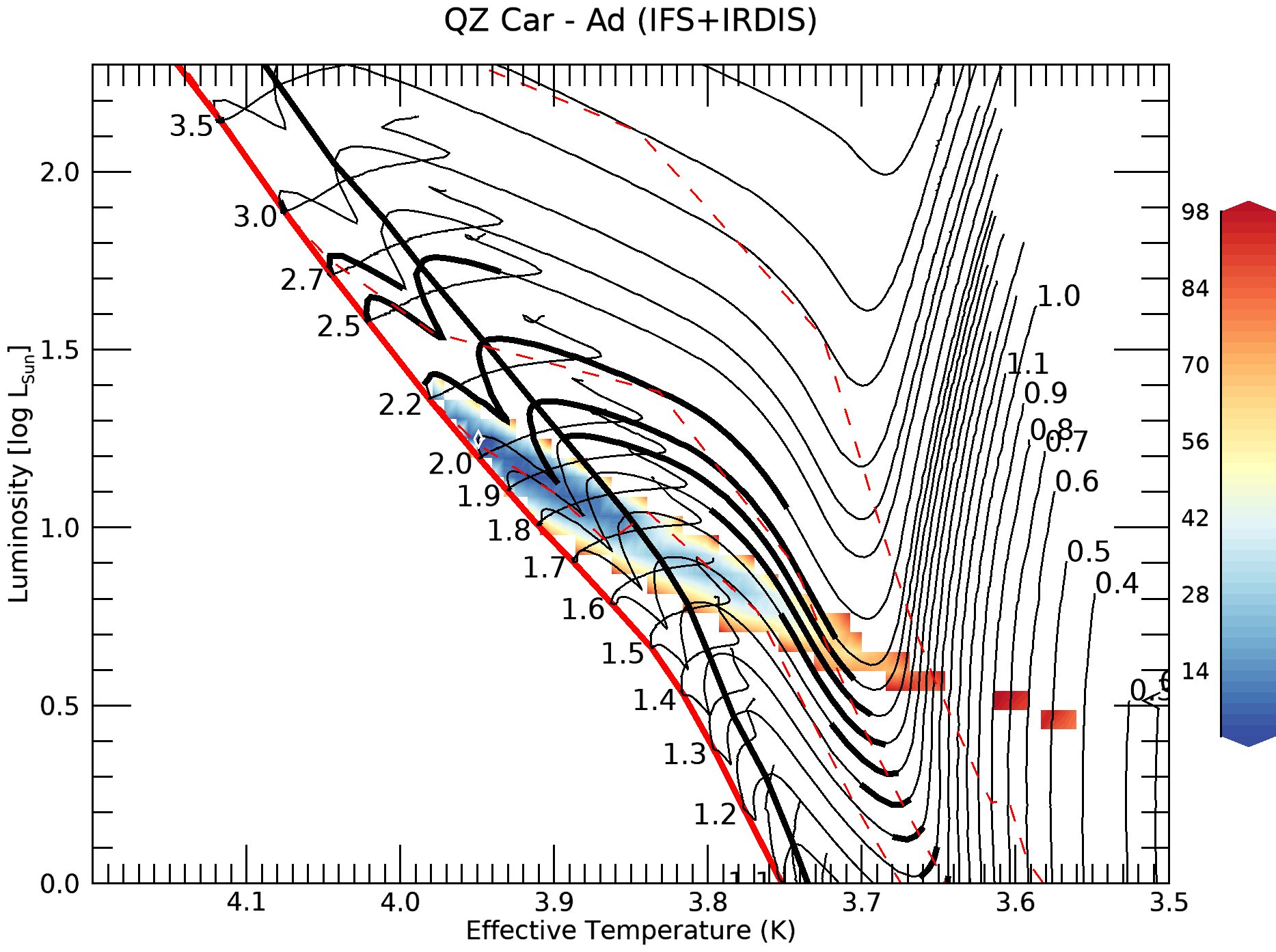}
     \caption{ $\chi^2$-surface resulting from ATLAS model fit to Ad's IFS+IRDIS SED projected onto the HRD plane. The thick black and red lines are ZAMS and early-MS according to \citet{siess2000} definitions$^5$. Thin black lines are the evolutionary PMS tracks for stars with masses increasing from bottom to top from 1.1 to 3.5~M$_\odot$. The 4 to 8~Myr parts of the respective evolutionary tracks are displayed with a thicker line. Finally, the red dashed lines give, from top to bottom, the 1, 5 and 10~Myr isochrones. The best fit model is indicated with a white diamond. }
    \label{f: chi2}
  \end{figure}

  \subsection{Distance of QZ~Car} \label{s: gaia}

Knowledge of the distance of QZ~Car with respect to the observer is crucial to convert the angular separation in physical (projected) separations. In an attempt to improve on the distance of QZ Car, we retrieved its astrometric information from the Gaia DR2 catalogue \citep{gaia2016,gaia2018,Lindegren2018}, including positions, parallax, proper motion, their uncertainties and correlations. We used a galactic prior with a length scale of 2.5 kpc \citep{walborn12} and performed an MCMC fit to the distance and proper-motion vector as described by \citet{bailerjones17}. We obtained a distance of $1.17^{+0.16}_{-0.13}$~kpc. Similar results  are obtained with a flat prior. This is in contrast with prior estimates of about 2.3 kpc \citep{walborn1995,smith2006}. We also computed a spectral distance modulus yielding a distance of about 2.0 to 2.1 kpc. Given these discrepancies, we scrutinised further the Gaia measurements. We computed the quality of the astrometric fit the so-called RUWE indicator of \citet{Lindegren2018}, leading to RUWE = 2.54, which reveals a poor astrometric fit despite the reasonable relative uncertainty of 11\% on the parallax, and the 299 good astrometric measurements. This poor RUWE is likely caused by the multiplicity of the central object, which  biases the Gaia parallax. This issue could be resolved in the next Gaia data release, where binarity will be taken into account when deriving the parallax.

    In the remainder of the paper, we adopt 2.3 kpc as the reference distance to convert the measured angular separations to (projected) physical distances. The results are given in Table~\ref{table:2}. We note that the only results in this work that depend on the adopted distance are the projected physical separations, so that there is no impact on the presented results if the adopted distance to QZ Car turned out to be incorrect.

  \begin{figure*}
  \centering
\includegraphics[width=.45\hsize]{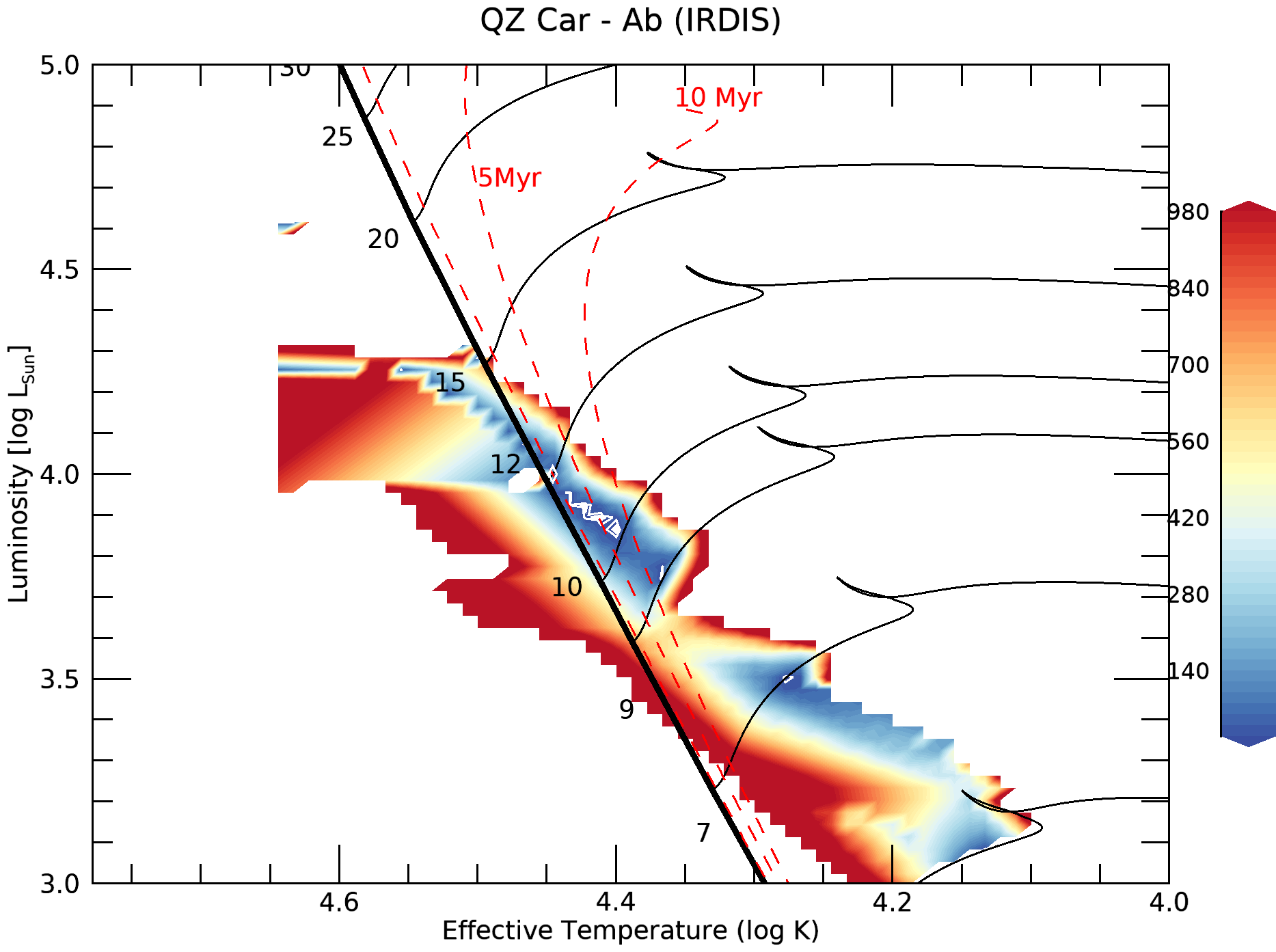}
  \includegraphics[width=.45\hsize]{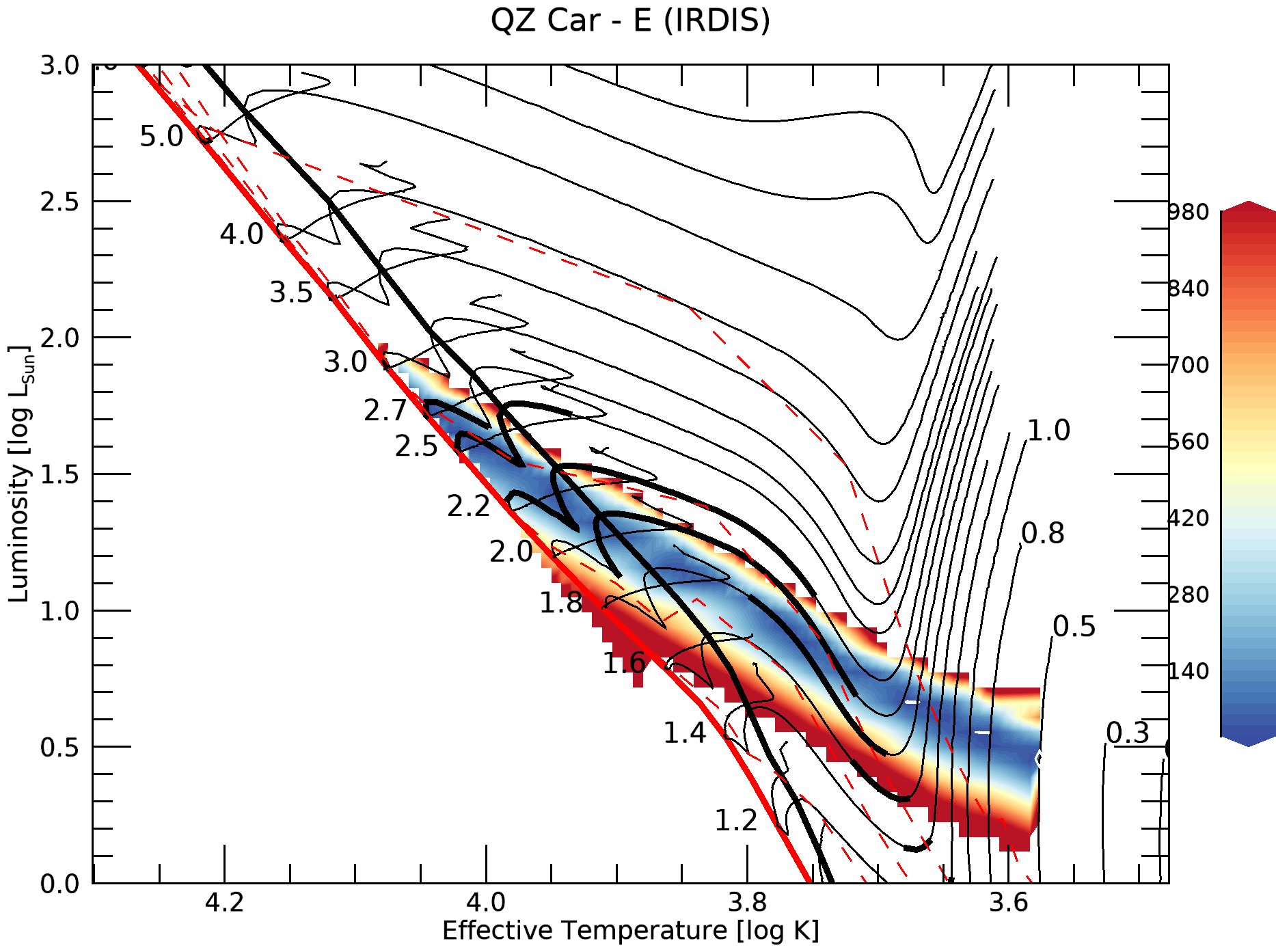}
     \caption{Same as Fig.~\ref{f: chi2} for the IRDIS sources Ab and E. }
    \label{f: chi2_irdis1}
  \end{figure*}

\section{Results and Discussion} \label{s: results}

  \subsection{Probabilities of spurious association} \label{s: Pspur}

    While over a dozen faint sources are clearly resolved around QZ~Car in Fig.~\ref{f: fov}, additional information such as common proper motion would be needed to confirm their physical connection to QZ~Car. Unfortunately, we only have one observation of QZ~Car so far and the closest object to QZ~Car that was also detected by Gaia is at 7.2", i.e. outside the IRDIS field-of-view. In the absence of such information, we resorted to a statistical argument. To this aim, we define
 the probability of spurious association ($P_\mathrm{spur}(\rho_i|\Sigma(K_i))$) as the probability that at least one source is found by chance at a separation $\rho$ equal or closer to QZ~Car than that of the companion $i$  ($\rho \le \rho_i$) given the local source density $\Sigma$ of stars at least as bright as $i$ ($K \le K_i$)\footnote{The definition of the probability of spurious association is modified compared to \citet{sana2014} in the sense that the present definition is an actual probability while the formula of \citet{sana2014} gives the expected number of companions within a given an angular separation and minimum brightness resulting from chance alignment from the local surface density of sources at least as bright.}.

 A query of the VISTA Carina Nebula catalogue \citep{Preibisch2014} yielded $N_\mathrm{obj} = 1864$ within a $r=2$\arcmin\ radius around QZ~Car down to magnitudes of $K_\mathrm{s} \approx 19$. To compute $P_\mathrm{spur}$, we first estimate the local source density $\Sigma(K_i)=N_\mathrm{obj}(K \le K_i)/(\pi r^2)$ of objects at least as bright as the companion $i$. We then use a Monte Carlo approach and randomly generate 10,000 populations of $N_\mathrm{obj}(K \le K_i)$ stars uniformly distributed in $\pi r^2$. The probability of spurious association is finally obtained as the fraction of populations in which at least one star is to be found at $\rho \le \rho_i$.

This simple exercise confirms the very low  probability of spurious association ($P_\mathrm{spur}<0.02$) -- hence the high confidence of physical association -- of companions Ab, Ad and E while the presence of most of the fainter 'S' sources are best explained by chance alignment given the overall surface density of sources in QZ~Car's surroundings.

In addition to computing the probability of spurious associations, we ran the Besan\c{c}on  model of the Galaxy \citep{robin2003} in the direction of QZ~Car. The model predicts about 20 stars with $K$-band magnitude brighter than 19~mag in an area corresponding  to our IRDIS field of view. All of them have $K> 15.5$ and the vast majority (18/21) are background stars (distance $>$ 3~kpc). To the first order, this is compatible with the properties of the S sources in the IRDIS field of view and provides an additional argument to consider (most of) them as chance alignments. It also supports the fact that Ad, Ab and E are physically connected as the Besan\c{c}on model cannot explain the presence of such bright objects around QZ~Car. Sources S1 to S6 have $0.25<P_\mathrm{spur} < 0.50$, so that, statistically, some of these could still be physically linked.
In summary, and accounting for the four inner objects, QZ~Car  consists of  seven likely physical companions within an $\approx$ 2\farcs5-radius, as well as additional  three to four fainter candidate companions within a similar angular separation.

Confirmation of common proper motion and characterization of orbital motion are of course crucial to definitely prove any physical association. For the closest companion, Ad, given the precision of our astrometry (see Table~\ref{table:2}) and the proper motion of the central star \citep{gaiadr2}, one should be able to prove common proper motion and measure a significant orbital rotation \citep[excluding contamination by high proper motion background objects, see][]{nielsen2017} with observations separated by 1 and 7 yr, respectively (assuming a circular orbit).

  \subsection{Spectral modelling of Ad} \label{s: SED}

    The flux calibrated low-resolution IFS+IRDIS spectrum of Ad provides us with information about its spectral energy distribution (SED). Here we attempt to use that information to constrain the stellar parameters of the Ad companion for the first time.

The uncalibrated IFS spectrum is mostly flat, except for two broad absorption features at 1120 nm and 1372 nm (also seen on the calibrated spectrum on Fig.~\ref{f: sed}). These correspond to the expected location of Earth telluric bands. To check this, we adjusted a synthetic telluric spectrum to the Ad data using MolecFit \citep{smette2015,kausch2015}.  We used the atmospheric conditions at the time of the observations and only consider telluric lines resulting from water. Once corrected for the telluric bands, and aside from (unphysical) edge effects, the Ad spectrum is feature-less as can be expected from its very-low spectral resolution. The present exercise is useful to confirm the origin of the broad absorption in the IFS Ad spectrum and to verify spectral ranges that can in principle be well corrected from telluric absorption. The MolecFit telluric corrected spectrum is however not used in the following as the division by the spectrum of the central object actually take care of the telluric correction automatically.

    To constrain the stellar parameters from the spectrum, we used ATLAS9 LTE atmosphere models \citep[][]{castelli2004}. We associated an ATLAS9 model to each time step in the pre-main sequence (PMS) evolutionary tracks of \citet{siess2000} and quantitatively compared it to Ad's SED. For numerical reasons in the computation of the $\chi^2$-contour curves, we also interpolated the ATLAS9 grid along the $\log g$-axis.

    Each ATLAS9 model was converted into flux at the reference distance of 100~\rsun\ and integrated over the width of each IFS and IRDIS wavelength channels, allowing us to compute the corresponding $\chi^2$ accounting for the error-bars on Ad's SED. Adopting the MCMC errors, the best-fit $\chi^2$ has a reduced value of 0.1, suggesting that the error bars are heavily overestimated. Adopting the Simplex+MC values yields a best-fit reduced- $\chi^2$  of about 4.4 suggesting that the latter are underestimated by a factor of about two. In the following we rescale the Simplex+MC errors so that the best-fit reduced- $\chi^2$ is equal to unity. The obtained $\chi^2$-map is displayed in the Hertzsprung-Russell diagram (HRD) of Fig.~\ref{f: chi2}.

    As expected, our fit results in multiple possible combinations but the allowed mass remains limited in the range of 1.8 to 2.2~M$_\odot$. The best fit model is obtained for a 2.0~M$_\odot$ star with $T_\mathrm{eff}=8896$~K, $\log g=4.27$, $L=17.7$~L$_\odot$ and $R=1.72$~R$_\odot$. The best fit spectrum from ATLAS9 shows a nice fit with the measured spectrum of Ad in Fig.~\ref{f: sed}. However, the precision of the model retrieval is limited by two factors: the anticipated degeneracy between physical surface properties and evolutionary stage and the  density of the model grid. The first is clearly illustrated in Fig.~\ref{f: chi2} by the elongated $\chi^2$-valley in the HRD. Focusing on the location of the best-fit model, there are statistically significant differences in the goodness-of-fit between the 2.0~M$_\odot$ PMS-tracks and the neighbouring 1.9 and 2.2~M$_\odot$-tracks. All other things being kept equal, these translates into statistical uncertainties of the order of 300~K and 0.2~dex in  $T_\mathrm{eff}$ and $\log L/L_\odot$, respectively. Should we have adopted the MCMC errors, $T_\mathrm{eff}$ and $L/L_\odot$ values of
    7500 to 9600~K and 11 to 23~$L_\odot$ would have been obtained within a 68\% confidence interval, respectively.

According to our best-fit solution, the star's evolutionary stage seems to be somewhere between \citet{siess2000}'s ZAMS and early-main sequence\footnote{ZAMS: defined as the time, after deuterium burning, when the nuclear luminosity provides at least 99\%\ of the total stellar luminosity. Early MS: defined as the time, when the star settles on the main sequence after the CN cycle has reached its equilibrium (this only affects stars with $M>~1.2$~M$_\odot$).}. Its age of 9.7~Myr is probably in fair agreement with QZ~Car age estimates \citep{walker2017}. The best-fit ATLAS9 model is displayed in Fig.~\ref{f: chi2} and should correspond to stars of spectral type of A3 according to \citet{siess2000} calibrations.

  \begin{figure}
  \centering
  \includegraphics[width=\hsize]{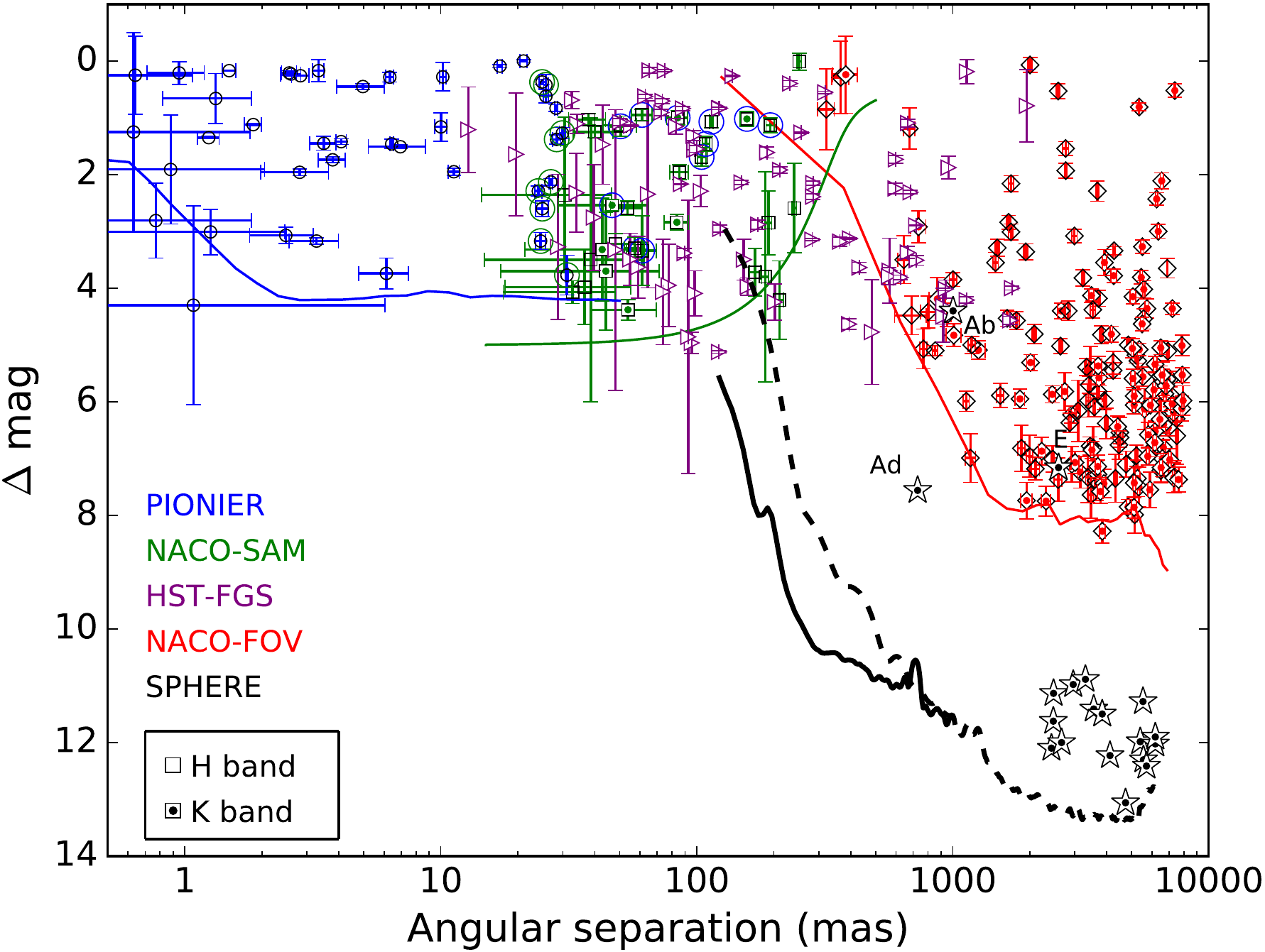}
     \caption{QZ~Car's sources detected with SPHERE (star symbol) overlaid on the SMaSH+ \citep{sana2014} and HST-FGS \citep{aldoretta2015} companion detections in the magnitude contrast vs.\ angular separation plane. The thick lines give the limiting contrast curves of the different instruments (see legend). }
    \label{f: smash}
  \end{figure}

  \subsection{Physical properties of the IRDIS companions} \label{s: irdis}

IRDIS observations only provide us with two independent wavelength channels, $K_1$ and $K_2$ with central wavelengths of 2.110 and 2.251~$\mu$m, respectively. While this is insufficient to constrain the shape of the SED, it provides an important anchor point to assess the objects absolute $K$-band magnitude {\it assuming} that the objects are located at the same distance and suffer from the same $K$-band reddening as QZ~Car.

As for the data of QZ Car Ad, we computed $\chi^2$ maps by comparing the $K_1$ and $K_2$ absolute fluxes of each companion sources to ATLAS9 LTE models. Good fits were obtained for most sources. The Ab companion seems to be more massive and we used the \citet{brott} main-sequence evolutionary tracks rather than the PMS-tracks from \citet{siess2000}.

The resulting $\chi^2$-maps, over-plotted in the HRD together with \citet{siess2000}'s PMS evolutionary tracks and isochrones are displayed in Fig.~\ref{f: chi2} and \ref{f: chi2_irdis1}. In this exercise, there is of course a degeneracy between the age and the mass.

 While more wavelength channels would be desirable, our results show that the companions Ab, Ad and E are compatible with the hypothesis of co-eval formation together with the central OB quadruple system in QZ~Car. Adopting the QZ~Car age range, i.e. 4 to 8 Myr, first-order constraints on the masses of the individual companions can be obtained. With only 4.4~mag contrast with QZ~Car central system, companion Ab is the most massive object and has a mass estimate of 10 to 12~M$_\odot$. The solution for E is fully degenerate with multiple minima on the $\chi^2$-map. We emphasise two of the extremer solutions: the best fit is a low mass (0.5 M$_\odot$) very young (2.5 Myr), cool ($T_{eff}=3770$K) and rather large ($R=3.7R_\odot$) PMS star. This solution is at the limit of the grid towards the low $T_{eff}$ so that it is possible cooler model may provide an even better fit. The other minimum is a 2.5 M$_\odot$, 5.5 Myr-old star with $T_{eff}=10$ kK which is also a valid solution within 3-sigma. The latter is compatible with co-evality.

We also performed this exercise for the faintest S sources, implicitly assuming that they are located in the Carina region. Under this hypothesis, sources S1 to S16 are compatible with being low-mass pre-main sequence stars with masses in the range of 0.3 to 0.5~M$_\odot$. But for S6 that could be as young as 10 to 15~Myr if its mass is on the low-side of the confidence interval, all the other sources seem to be older than 20~Myr. Alternatively, they are even younger, lower-mass PMS star falling outside of the ATLAS9 model-grid or they are foreground/background sources unconnected to the Carina region as discussed in Sect.~\ref{s: Pspur}.

  \subsection{Comparison with previous high-angular resolution surveys} \label{s: smash}

   Our results can be directly compared to previous high-angular resolution campaigns such as those provided by the SMaSH+ and HST-FGS surveys  \citep[Fig.~\ref{f: smash},][]{sana2014,aldoretta2015}. Clearly, by enabling the detection of sources at least 5~magnitudes fainter than previously possible, SPHERE opens a new discovery space to investigate the low-mass end of the companion mass function of massive stars.

   In this context, we note a clear clustering of the 'S' sources in Fig.~\ref{f: smash}, that are all located at angular separations larger than 2" (i.e., projected physical separation $>4.6\times 10^3$~au). Furthermore, there is a clear gap of 4- to 5-mag between the more massive, very likely physical and probably co-eval companions Ab, Ad and E and the rest of the 'S' sources, most of which are either older or unconnected, as suggested by the large spurious alignment probabilities that we derived.

    \begin{figure}
        \centering
        \includegraphics[width=1.05\hsize]{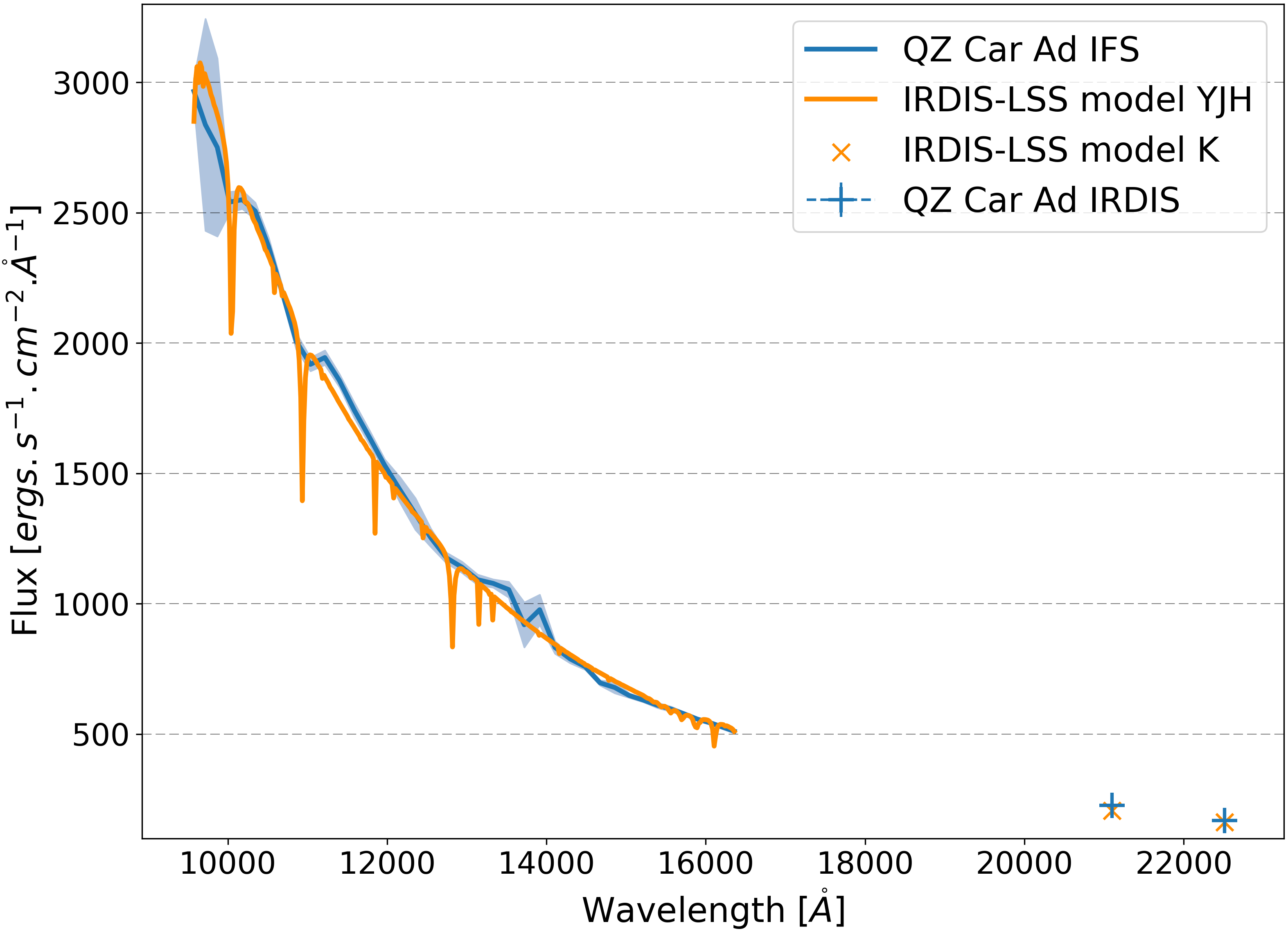}
        \caption{Spectrum of QZ Car Ad (\textit{blue}). Same figure as Fig.~\ref{f: sed} with the ATLAS9 \citep{castelli2004} models downgraded to the resolution of IRDIS-LSS (orange; $\lambda / \Delta \lambda=350$).}
        \label{f: LSS}
    \end{figure}

  \subsection{Future prospects} \label{s: future}

    The spectrum of QZ Car Ad that we obtained with IFS in the IRDIFS\_EXT mode has a spectral resolving power ($\lambda / \Delta \lambda$) of $\sim$50 only. At such low resolution, all spectral features but the telluric bands are smeared out (see Fig.~\ref{f: sed}). To better characterise the companion physical properties and age, a higher-resolution spectrum is desirable. However, seeing-limited spectrographs such as VLT/XSHOOTER will not be able to resolve the Ad companion. Very few AO-assisted spectrographs exist and, among those, almost none can deliver the require flux contrast. In Fig.~\ref{f: LSS}, we investigate the resolving power of the Long Slit Spectroscopic (LSS) mode of IRDIS, delivering a spectral resolving power of 350, which would provide a valuable improvement to the current IFS SED and help us to estimate the parameters of the companion with greater accuracy.

\section{Conclusions} \label{s:Ccl}

    We have presented the first SPHERE observations of QZ~Car, a known quadruple system in the Carina region. Using the IRDIFS\_EXT mode, we detected 19 sources in a 12"$\times$12" field-of-view; all but two (Ab and E) are newly detected. We used the high-contrast imaging software VIP used for planet detection, to characterise the detected sources. Three of our sources (Ab, Ad and E) are moderately bright, with $K$-band magnitude contrasts in the range of $\sim4$ to 7.5. The remaining sources have magnitude contrast from 10 to 13. Most of the latter can be explained by spurious alignment given the source number density around QZ~Car.

    We further determined the limiting contrast curves, showing that SPHERE detection capabilities can reach contrasts better than 9 mag at only 200~mas and better than 13 mag at angular separations larger 2". Our observations are sensitive to sub-solar mass companions over most of the angular separation range provided by SPHERE.

    Finally, we used the known distance to QZ~Car, a grid of ATLAS9 models and pre-main sequence evolutionary tracks to obtain a first estimate of the physical properties of the detected objects. This determination implicitly assumes that the companions are located at the same distance and suffer from the same NIR reddening than QZ~Car central system. We found masses across the entire mass range, from a fraction of a solar mass up to 12~M$_\odot$, including a $\sim2.0$~M$_\odot$ companion at a (projected) physical separation less than 1700~au. While there is a degeneracy in the physical parameter vs.\ age determination given the limited constraints, the three most massive, likely physical companions (Ab, Ad and E) can be fitted with ages of 4 to 9~Myr, i.e.\ their formation is potentially contemporaneous to that of the inner quadruple system making QZ~Car one of the highest order multiple system known.

    Future work can follow two directions. On the one hand, a better characterisation of the detected companions is desirable and will ultimately provide an independent age diagnostic. This will help to confirm physical connection of the companion through proper motions as well as  high-resolution spectroscopy and may be possible with the SPHERE Long Slit Spectrograph (LSS) for the brightest companions of QZ~Car.

    On the other hand, and based on the present results, it is clear that SPHERE is opening a new parameter space to investigate the presence and physical properties of faint companions within only a few 1000~au from massive stars. Performing similar observations of the entire sample of massive stars may allow us to investigate the outcome of the massive star formation process as well as to investigate the pairing mechanism of these faint companions.

\section*{Acknowledgements}

  This work is based on observations collected at the European Southern Observatory under programs ID 096.C-0510(A). We thank the SPHERE Data Centre, jointly operated by OSUG/IPAG (Grenoble), PYTHEAS/LAM/CeSAM (Marseille), OCA/Lagrange (Nice) and Observatoire de Paris/LESIA (Paris) and supported by a grant from Labex OSUG@2020 (Investissements d'avenir a ANR10 LABX56). We especially thank P. Delorme and E. Lagadec (SPHERE Data Centre) for their help during the data reduction process.

  We acknowledge support from the FWO-Odysseus program under project G0F8H6N. This project has further received funding from the European Research Council under European Union's Horizon 2020 research programme (grant agreement No 772225) and under the European Union's Seventh Framework Program (ERC Grant Agreement n. 337569).
  LAA acknowledges financial support by Coordena\c{c}\~{a}o de Aperfei\c{c}oamento de Pessoal de Nível Superior and Funda\c{c}\~{a}o de Amparo \`{a} Pesquisa do Estado de S\~{a}o Paulo.
  This work has made use of data from the European Space Agency (ESA) mission
{\it Gaia} (\url{https://www.cosmos.esa.int/gaia}), processed by the {\it Gaia}
Data Processing and Analysis Consortium (DPAC,
\url{https://www.cosmos.esa.int/web/gaia/dpac/consortium}). Funding for the DPAC
has been provided by national institutions, in particular the institutions
participating in the {\it Gaia} Multilateral Agreement. JDR acknowledges the
BELgian federal Science Policy Office (BELSPO) through PRODEX grants Gaia
and PLATO.
VC acknowledges funding from the Australian Research Council via DP180104235.

  \textit{Facilities:} VLT UT3 (SPHERE)

%
\pagebreak

\clearpage
\newpage

\begin{appendix} 
\section{Spectrum of QZ Car Ad} \label{a: Adspec}
\begin{table*}
\caption{Contrast and flux calibrated spectrum values for Ad and model spectra of QZ Car components}
\centering
\begin{tabular}{c c c c c c c }     
\hline\hline
Wavelength & Contrast spectrum  & Flux calibrated spectrum  & Aa1 & Aa2  & Ac1 & Ac2 \\
($\mathring{A}$) & ($10^{-4}$) & \multicolumn{5}{c}{[$10^{3}$\,ergs\,s$^{-1}$\,cm$^{-2}$\,$\mathring{A}^{-1}$]} \\
\hline
   957 & $6.20\pm0.48$ & $2.96\pm0.23$ & 2571.51 & 25.85 & 1123.46 & 336.55\\
   972 & $6.27\pm0.97$ & $2.83\pm0.44$ & 2430.00 & 24.51 & 1060.55 & 318.00\\
   987 & $6.44\pm0.85$ & $2.75\pm0.36$ & 2293.54 & 23.10 & 999.97 & 300.05\\
   1002 & $6.31\pm0.29$ & $2.54\pm0.12$ & 2164.20 & 21.93 & 942.70 & 282.86\\
   1018 & $6.73\pm0.31$ & $2.55\pm0.12$ & 2039.48 & 20.71 & 887.51 & 266.30\\
   1034 & $7.02\pm0.33$ & $2.50\pm0.12$ & 1920.56 & 19.55 & 834.97 & 250.57\\
   1051 & $7.06\pm0.27$ & $2.37\pm0.09$ & 1807.16 & 18.43 & 784.91 & 235.61\\
   1068 & $6.95\pm0.25$ & $2.19\pm0.08$ & 1699.00 & 17.37 & 737.20 & 221.35\\
   1085 & $6.76\pm0.30$ & $2.01\pm0.09$ & 1598.02 & 16.36 & 692.90 & 207.96\\
   1103 & $6.88\pm0.37$ & $1.92\pm0.10$ & 1501.89 & 15.41 & 650.65 & 195.31\\
   1122 & $7.42\pm0.38$ & $1.94\pm0.10$ & 1410.82 & 14.50 & 610.65 & 183.32\\
   1140 & $7.53\pm0.32$ & $1.85\pm0.08$ & 1326.19 & 13.65 & 573.49 & 172.18\\
   1159 & $7.52\pm0.30$ & $1.74\pm0.07$ & 1247.08 & 12.84 & 538.78 & 161.75\\
   1178 & $7.53\pm0.28$ & $1.63\pm0.06$ & 1172.49 & 12.08 & 506.07 & 151.93\\
   1197 & $7.48\pm0.27$ & $1.53\pm0.05$ & 1102.32 & 11.37 & 475.38 & 142.71\\
   1216 & $7.46\pm0.34$ & $1.43\pm0.07$ & 1036.57 & 10.71 & 447.04 & 134.16\\
   1235 & $7.41\pm0.41$ & $1.34\pm0.07$ & 974.82 & 10.09 & 420.42 & 126.12\\
   1255 & $7.34\pm0.31$ & $1.25\pm0.05$ & 918.27 & 9.51 & 395.74 & 118.70\\
   1274 & $7.32\pm0.25$ & $1.17\pm0.04$ & 866.50 & 8.97 & 372.89 & 111.85\\
   1294 & $7.53\pm0.30$ & $1.14\pm0.04$ & 817.91 & 8.46 & 351.46 & 105.43\\
   1313 & $7.63\pm0.26$ & $1.09\pm0.04$ & 772.53 & 7.99 & 331.65 & 99.48\\
   1333 & $7.97\pm0.44$ & $1.07\pm0.04$ & 730.03 & 7.55 & 313.19 & 93.93\\
   1352 & $8.24\pm0.95$ & $1.05\pm0.06$ & 690.47 & 7.14 & 295.98 & 88.758\\
   1372 & $7.59\pm1.22$ & $0.92\pm0.12$ & 653.73 & 6.76 & 280.05 & 83.98\\
   1391 & $8.51\pm1.29$ & $0.97\pm0.15$ & 619.62 & 6.41 & 265.25 & 79.54\\
   1411 & $7.64\pm1.24$ & $0.83\pm0.15$ & 591.79 & 6.25 & 253.05 & 75.98\\
   1430 & $7.62\pm0.59$ & $0.79\pm0.13$ & 562.40 & 5.94 & 240.32 & 72.13\\
   1449 & $7.71\pm0.44$ & $0.76\pm0.06$ & 534.93 & 5.65 & 228.43 & 68.55\\
   1467 & $7.39\pm0.41$ & $0.69\pm0.04$ & 509.27 & 5.37 & 217.33 & 65.20\\
   1486 & $7.55\pm0.42$ & $0.67\pm0.04$ & 485.46 & 5.12 & 207.04 & 62.10\\
   1503 & $7.55\pm0.32$ & $0.64\pm0.04$ & 463.61 & 4.88 & 197.60 & 59.26\\
   1522 & $7.69\pm0.31$ & $0.63\pm0.03$ & 443.21 & 4.67 & 188.79 & 56.62\\
   1539 & $7.79\pm0.27$ & $0.61\pm0.02$ & 424.17 & 4.46 & 180.57 & 54.15\\
   1556 & $7.93\pm0.32$ & $0.59\pm0.02$ & 406.42 & 4.27 & 172.92 & 51.85\\
   1573 & $8.00\pm0.34$ & $0.58\pm0.02$ & 393.71 & 4.14 & 167.47 & 50.21\\
   1589 & $8.04\pm0.32$ & $0.56\pm0.02$ & 385.65 & 4.05 & 164.04 & 49.18\\
   1605 & $8.15\pm0.41$ & $0.54\pm0.02$ & 378.05 & 3.97 & 160.81 & 48.21\\
   1621 & $8.20\pm0.49$ & $0.53\pm0.03$ & 370.89 & 3.90 & 157.76 & 47.30\\
   1636 & $8.23\pm0.74$ & $0.51\pm0.04$ & 364.15 & 3.83 & 154.89 & 46.44 \\
   2110 & $9.72\pm0.18$ & $0.22\pm0.04$ & 141.87 & 1.54 & 61.69 & 17.70 \\
   2251 & $9.20\pm0.77$ & $0.16\pm0.13$ & 110.05 & 1.20 & 47.84 & 13.66 \\
\hline
\end{tabular}
\end{table*}

\pagebreak
\clearpage
\section{Comparison between parameter estimation techniques: Simplex, MCMC \& PSF-fitting} \label{a: comparison}

  \begin{figure*}
  \centering
\includegraphics[width=.45\hsize]{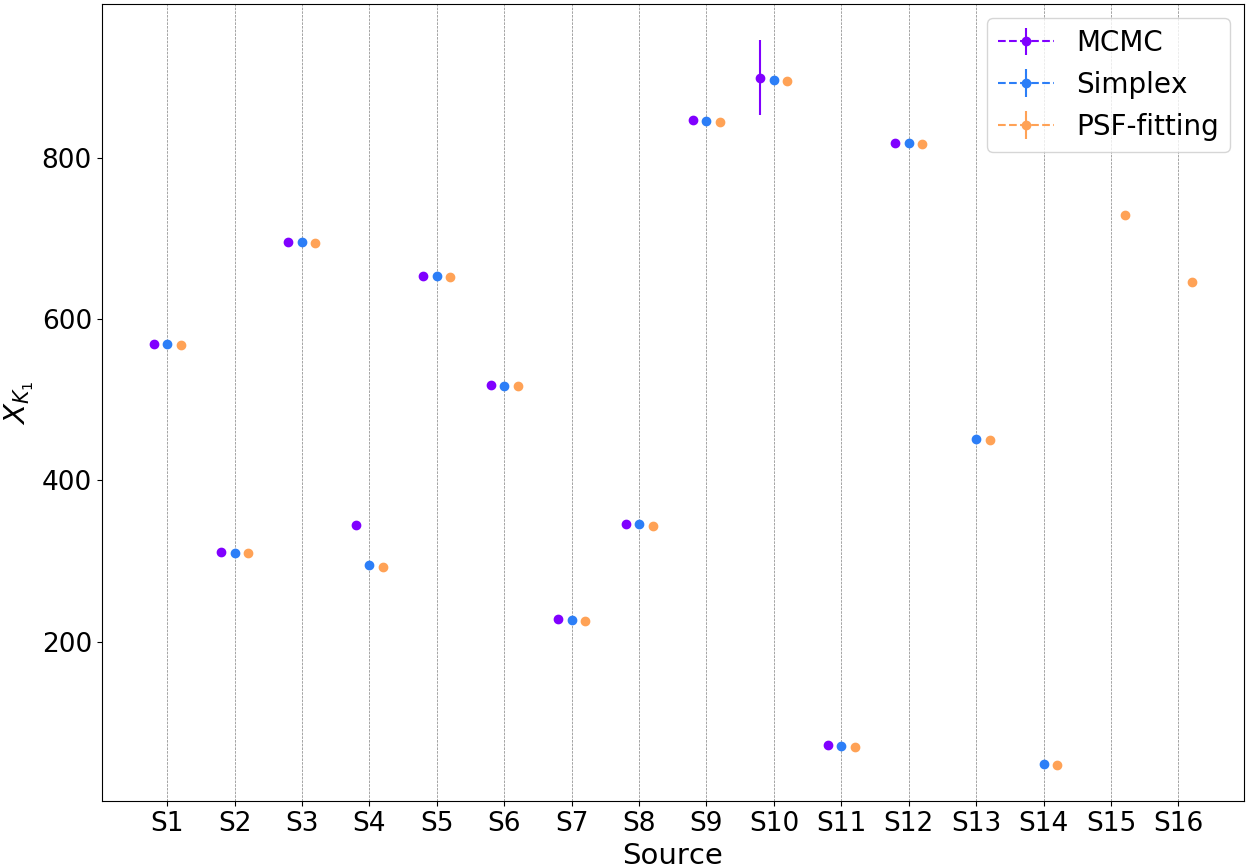}
  \includegraphics[width=.45\hsize]{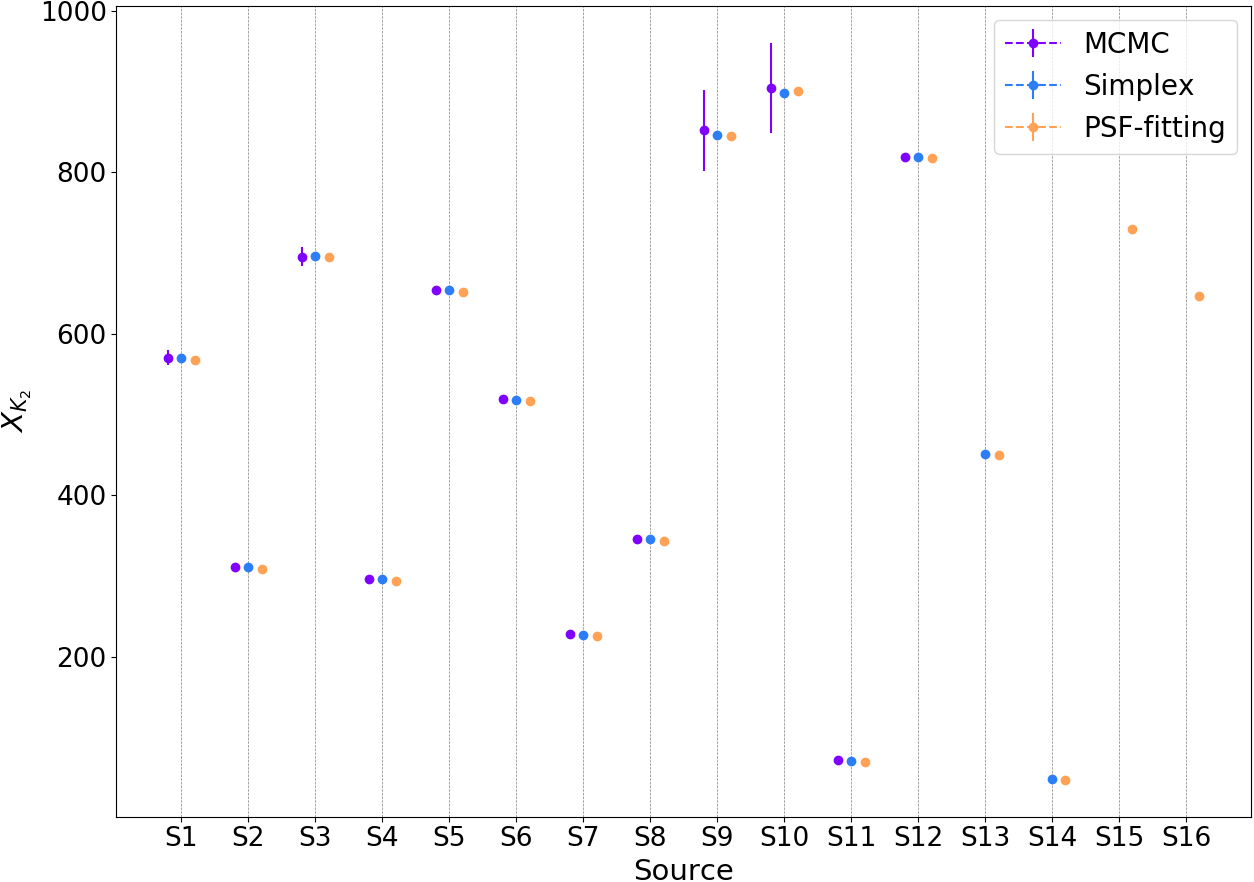}
     \caption{Comparison of the $X$ coordinates and associated errors. Most errors for Simplex and PSF fitting are small and not seen on these plots. \textit{Left:} $K_1$. \textit{Right:} $K_2$}
     \label{f: comparison_x}

  \end{figure*}

  \begin{figure*}
  \centering
\includegraphics[width=.45\hsize]{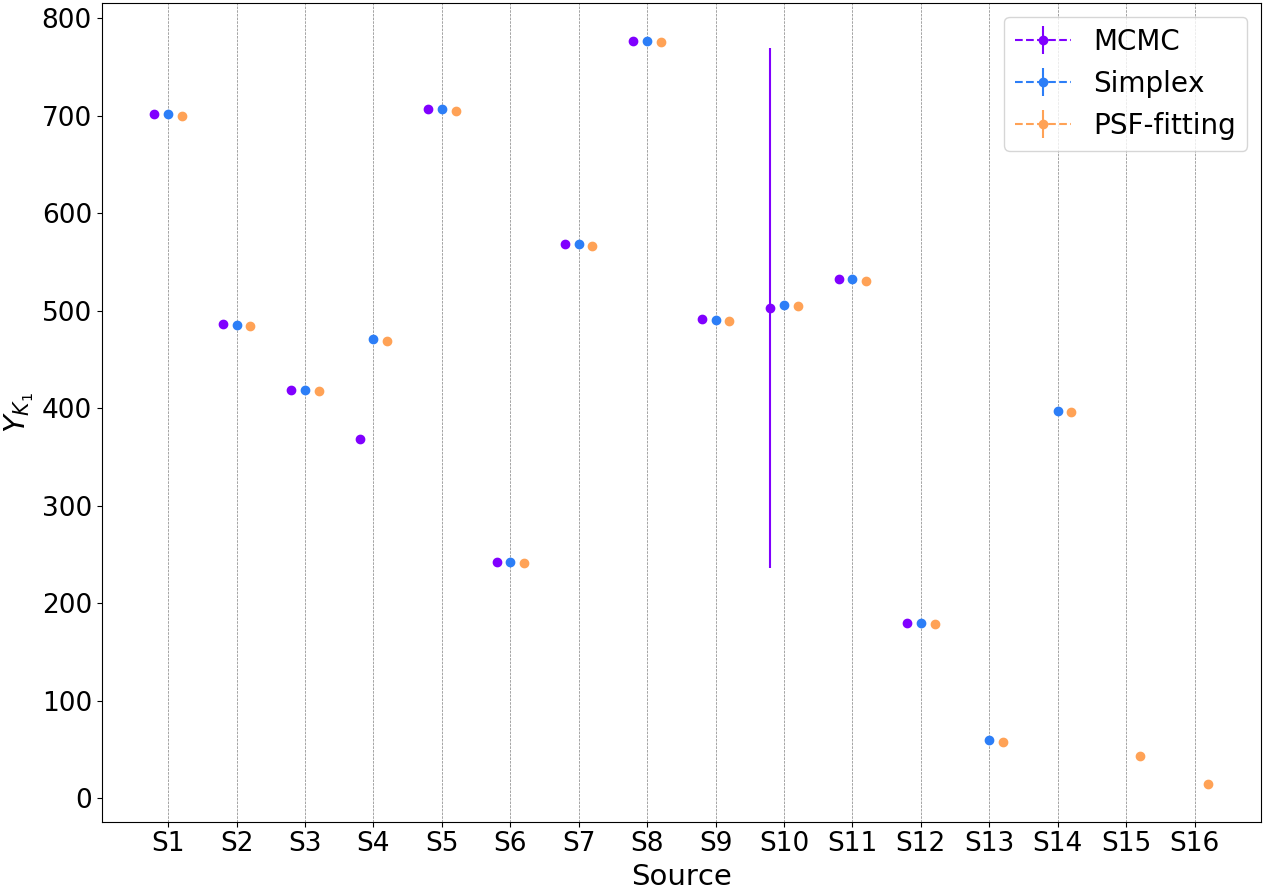}
  \includegraphics[width=.45\hsize]{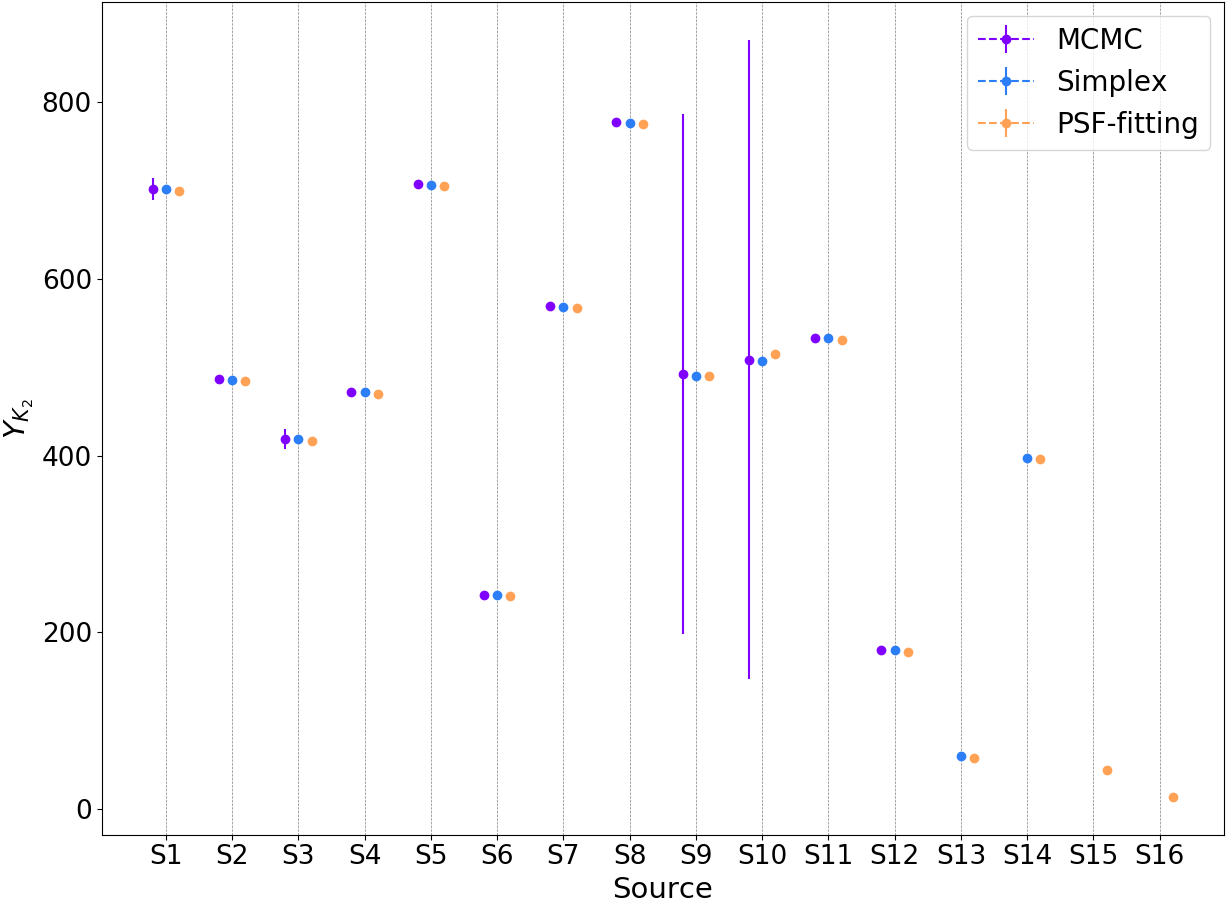}
     \caption{Comparison of the $Y$ coordinates and associated errors. Most errors for Simplex and PSF fitting are small and not seen on these plots. \textit{Left:} $K_1$. \textit{Right:} $K_2$}

     \label{f: comparison_y}
  \end{figure*}

%





\end{appendix}

\end{document}